\documentclass[conference]{IEEEtran}
\IEEEoverridecommandlockouts

\usepackage{cite}
\usepackage{amsmath,amssymb,amsfonts}
\usepackage{graphicx}
\usepackage{textcomp}
\usepackage{xcolor}
\usepackage{fancyhdr} 
\usepackage{adjustbox}

\usepackage{color}
\usepackage{multirow}
\usepackage{booktabs}
\usepackage{caption}   
\usepackage{subfigure} 
\usepackage{comment}
\usepackage{mathtools}
\usepackage{mathrsfs}
\usepackage{algorithm}
\usepackage{algpseudocode}
\usepackage{amsmath}
\usepackage{siunitx}

\newtheorem{definition}{Definition}

\newtheorem{assumption}{Assumption}[section]
\def\BibTeX{{\rm B\kern-.05em{\sc i\kern-.025em b}\kern-.08em
    T\kern-.1667em\lower.7ex\hbox{E}\kern-.125emX}}

\DeclareRobustCommand*{\IEEEauthorrefmark}[1]{%
    \raisebox{0pt}[0pt][0pt]{\textsuperscript{\footnotesize\ensuremath{#1}}}}

\begin{document}

\title{ReliCD: A Reliable Cognitive Diagnosis Framework with Confidence Awareness}

\author{
\IEEEauthorblockN{
Yunfei Zhang\IEEEauthorrefmark{1,*} \thanks{* Yunfei Zhang and Chuan Qin contribute equally to this research.},
Chuan Qin\IEEEauthorrefmark{2,*},
Dazhong Shen\IEEEauthorrefmark{3},
Haiping Ma\IEEEauthorrefmark{1,\dagger}\thanks{$\dagger$ Haiping Ma and Hengshu Zhu are corresponding authors.},
Le Zhang\IEEEauthorrefmark{4},
Xingyi Zhang\IEEEauthorrefmark{5},
Hengshu Zhu\IEEEauthorrefmark{2,\dagger}
}
\IEEEauthorblockA{
\IEEEauthorrefmark{1}Department of Information Materials and Intelligent Sensing Laboratory of Anhui Province, Institutes of Physical Science and  \\
Information Technology, Anhui University, China, 
\{a2946634883@163.com, hpma@ahu.edu.cn\}}
\IEEEauthorblockA{
\IEEEauthorrefmark{2}Career Science Lab, BOSS Zhipin, China, 
\{chuanqin0426, zhuhengshu\}@gmail.com}
\IEEEauthorblockA{
\IEEEauthorrefmark{3}Shanghai Artificial Intelligence Laboratory, China, 
dazh.shen@gmail.com}
\IEEEauthorblockA{\IEEEauthorrefmark{4}Business Intelligence Lab, Baidu Inc, China, 
zhangle0202@gmail.com}
\IEEEauthorblockA{\IEEEauthorrefmark{5}School of Computer Science and Technology, Anhui University, China, xyzhanghust@gmail.com}
}

\maketitle

\begin{abstract}
During the past few decades, cognitive diagnostics modeling has attracted increasing attention in computational education communities, which is capable of quantifying the learning status and knowledge mastery levels of students. Indeed, the recent advances in neural networks have greatly enhanced the performance of traditional cognitive diagnosis models through learning the deep representations of students and exercises. 
Nevertheless, existing approaches often suffer from the issue of overconfidence in predicting students' mastery levels, which is primarily caused by the unavoidable noise and sparsity in realistic student-exercise interaction data, severely hindering the educational application of diagnostic feedback. 
To address this, in this paper, we propose a novel \textbf{Reli}able \textbf{C}ognitive \textbf{D}iagnosis (ReliCD) framework,  which can quantify the confidence of the diagnosis feedback and is flexible for different cognitive diagnostic functions. 
Specifically, we first propose a Bayesian method to explicitly estimate the state uncertainty of different knowledge concepts for students, which enables the confidence quantification of diagnostic feedback. 
In particular, to account for potential differences, we suggest modeling individual prior distributions for the latent variables of different ability concepts using a pre-trained model.
Additionally, we introduce a logical hypothesis for ranking confidence levels. Along this line, we design a novel calibration loss to optimize the confidence parameters by modeling the process of student performance prediction.  
Finally, extensive experiments on four real-world datasets clearly demonstrate the effectiveness of our ReliCD framework.

\end{abstract}


\section{Introduction}
Cognitive diagnosis, as an essential component of computer-aided education, has garnered increasing attention over the past decades~\cite{leighton2007cognitive,NCD,RCD}. 
The primary objective of cognitive diagnostics modeling is to quantitatively assess students' learning status and knowledge mastery levels, providing valuable formative feedback~\cite{leighton2007cognitive,NCD}. 
Indeed, relevant studies have enabled a wide range of downstream educational applications, such as course recommendations~\cite{course}, student assessment~\cite{khajah2014integrating}, and computerized adaptive testing~\cite{ma2023novel}. 
As shown in Figure 1(a), given the answering records of student Lano concerning  a series of exercises, the cognitive diagnosis model can automatically estimate her mastery levels of various knowledge concepts.

\begin{figure}[!t]
\centering
\includegraphics[width=0.42 \textwidth]{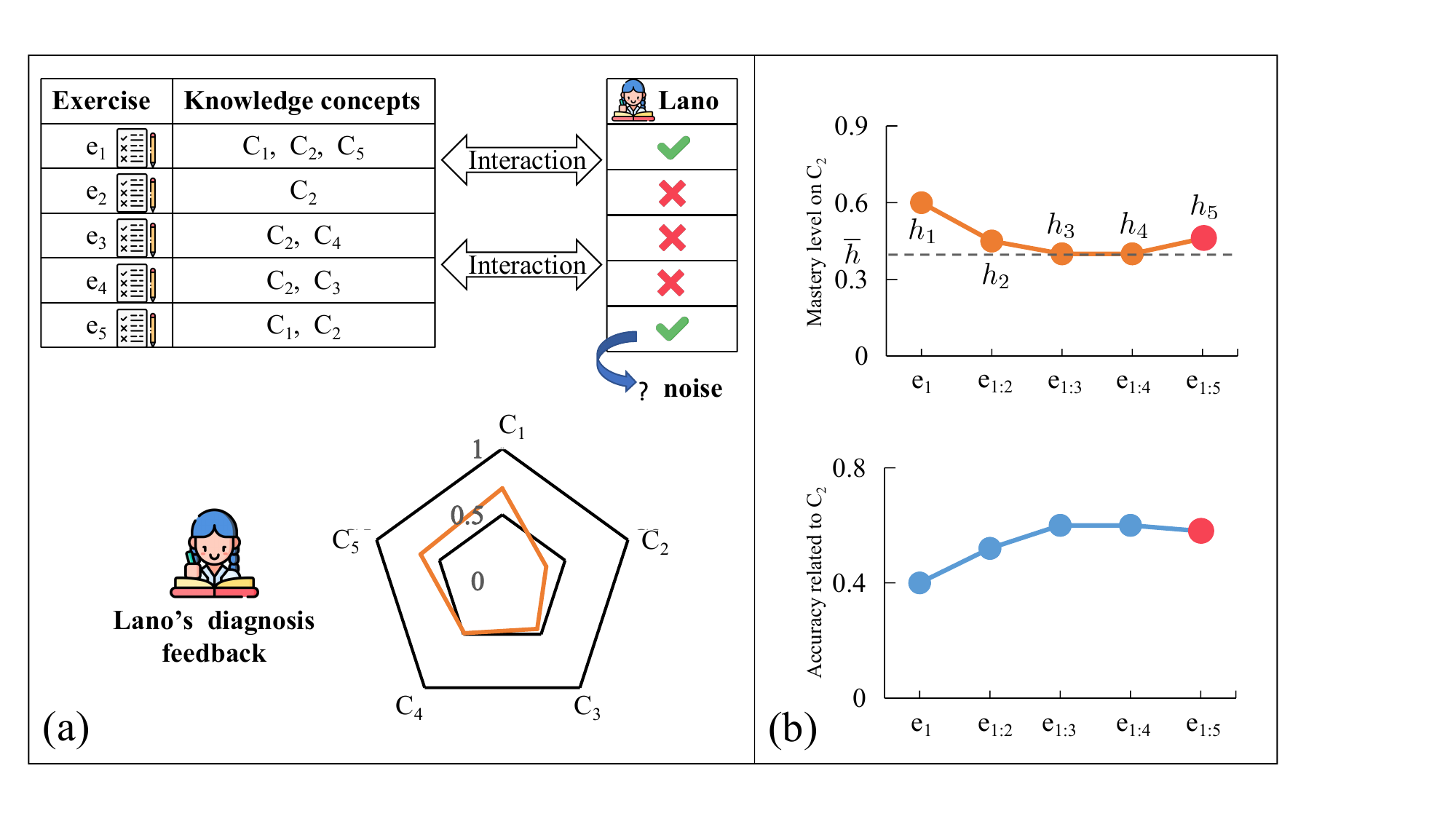} 

\caption{(a)~An example of cognitive diagnosis; (b)~the predicted Lano's diagnostic feedback on concept $C_2$ with different interaction data and the corresponding accuracy of her performance prediction on all the exercises related to the concept $C_2$ in the test set, where $e_{1:j}$ denotes the exercises set $\{e_1, e_2, ..., e_j\}$ and $\bar{h}$ indicates Lano's actual ability state on $C_2$.}
\label{fig1}
\end{figure}

In the literature, traditional cognitive diagnosis models~(CDMs) utilize different linear psychometric functions to measure students' learning status  by modeling the process of student performance prediction, such as Deterministic Inputs, Noisy ``And'' gate~(DINA)~\cite{DINA}, Item Response Theory (IRT)~\cite{IRT}. Recently, with the rapid development of deep learning techniques, several neural-based cognitive diagnostic methods have been proposed to enhance diagnostic performance.
For instance, the neural cognitive diagnosis framework (NCD) utilizes neural networks to model students-exercise interactions, in order to uncover deeper features of both students and exercises~\cite{NCD}. 
Moreover, the flexibility of neural model design has enabled researchers to incorporate additional information, such as concept dependency maps~\cite{RCD} and student profiling~\cite{zhou2021modeling}, to further improve the effectiveness and interpretability of the models.

Previous studies have evaluated the effectiveness of cognitive diagnostic models by calculating the accuracy of student performance prediction, but they have not measured the reliability of diagnostic feedback. Meanwhile, due to the presence of noise and sparsity in student-exercise interaction data, existing approaches lead to the potential overconfidence in students’ mastery prediction, severely reducing the reliability of real-time  diagnostic feedback in practical online education systems. More specifically, as illustrated in Figure~1(b), when Lano interacted with each exercise (i.e., from \textsl{$e_1$} to \textsl{$e_5$}), we present the cognitive diagnosis model's results regarding her mastery of knowledge concept \textsl{$C_2$} and the accuracy of her performance prediction on all the exercises related to the concept \textsl{$C_2$} in the test set. We found that due to the noise present in the interaction data (i.e., $<$Lano, \textsl{$e_5$}, \textcolor{green}{$\checkmark$}$>$), the mastery of \textsl{$C_2$} at $h_5$ deviates from the actual state $\bar{h}$. It indicates that we cannot trust the diagnostic feedback in a monotonic manner with the increase in interaction. Furthermore, traditional evaluation metrics like accuracy are non-smooth functions, which can result in the same evaluation outcome despite different diagnostic feedback. Additionally, these indicators are often not available in real-time during the diagnostic process in practical use. Consequently, an ideal cognitive diagnosis model should be able to provide both accurate diagnostic feedback and indications of its reliability.


To this end, in this paper, we propose a novel reliable cognitive diagnosis framework, namely ReliCD. 
To the best of our knowledge, this is the first one to quantify the confidence of the diagnosis feedback and is flexible for different cognitive diagnostic functions. 
Specifically, we first propose a Bayesian method for explicitly estimating the uncertainty of students'  states for various knowledge concepts with Gaussian latent variables, where the mean parameter represents the average ability status and the variance enables the quantification of diagnostic feedback’s confidence.
In particular, due to the potential difference, we model the individual prior distribution for the latent variables of different ability concepts with a pre-trained model.
Then, we introduce a logical hypothesis for ranking confidence levels and present a novel calibration loss to optimize the parameters in determining diagnostic feedback’s confidence through modeling the process of student performance prediction.  
Finally, extensive experiments on four real-world datasets  demonstrate the effectiveness and flexibility of our ReliCD.

\section{Related Work}
Generally, the related work in this paper can be grouped into two categories: cognitive diagnosis and confidence estimation.

\subsection{Cognitive Diagnosis}
The main task of cognitive diagnosis is to use students' responses to exercises for diagnosing students' ability state. Over the past decades, experts in related educational
psychology fields have proposed many cognitive diagnostic models. The two most classic ones are IRT~\cite{IRT} and DINA~\cite{DINA}. In IRT, \emph{Embretson et al.} represented students' ability state as a one-dimensional  and continuous scalar. And a logistic function is used to predict the probability that the student eventually responds correctly to the exercise. Later, some researchers improved upon IRT and proposed MIRT~\cite{MIRT} by extending the ability state of students to multi-dimensional vectors. Different from IRT, DINA uses a binary vector to model 
 the student's ability state with each dimension's value representing his/her mastery of relevant knowledge concepts. There are two possible values on each dimension, 1~(mastered) or 0~(not mastered).
Furthermore, \emph{Jimmy De La Torre} believed that DINA itself has strong assumptions and constraints, which do not conform to the actual situation. 
Along this line, they proposed a generalized DINA (G-DINA)~\cite{GDINA} to improve the diagnostic performance by weakening these constraints.

In recent years, neural-based cognitive diagnosis models have achieved state-of-the-art prediction performance, benefiting from the successful application of neural networks in various fields, including recommendation systems~\cite{recom_1}, knowledge tracing~\cite{ma2022reconciling}, and computer vision~\cite{cv_1}.
These works can be mainly divided into two aspects. The first aspect focuses on designing diagnostic functions that leverage the power of neural networks to capture complex and non-linear interactions between students and exercises, such as NCD~\cite{NCD}. 
The second is to use neural networks to enrich the representation of students and exercises by considering more additional information~(e.g., the exercise text information, the relationship between knowledge concepts).  For example, deep IRT~(DIRT)~\cite{DIRT} uses the semantic information of the exercise text to enrich the parameter representation of the traditional IRT. Educational context-aware cognitive diagnosis~(ECD)~\cite{ECD} was proposed by incorporating the student's educational background into the modeling of student knowledge status. \emph{Gao et al.}~\cite{RCD} proposed the relation map-driven cognitive diagnosis~(RCD) framework by exploiting  the prerequisite relation and similarity relationship of  knowledge concepts.  
\emph{Ma~et~al.}~\cite{PAKP} proposed a prerequisite attention model~(PAKP) for knowledge proficiency diagnosis of students by considering the prerequisite relationship of knowledge concepts and learning the influence weights of predecessor knowledge concepts on successor knowledge concepts. Furthermore, \emph{Li et al.}~\cite{HCDF} proposed a novel CDM, namely HCDF, to enhance diagnostic performance by modeling the hierarchical relationship between knowledge concepts.  

The majority of existing studies primarily concentrate on enhancing the accuracy of student performance prediction. However, there has been a notable lack of comprehensive investigation into the aspect of reliability in diagnostic feedback. In this paper, we introduce a novel approach for reliable cognitive diagnosis, which is the first to quantitatively assess the reliability of diagnostic feedback.

\subsection{Confidence Estimation}
Confidence estimation has been incorporated within the machine learning community in some specific areas including autonomous driving~\cite{drive_1}, medical applications~\cite{medical_1},  
and career mobility analysis~\cite{zha2023career, qin2023comprehensive}, so as to provide insight into the reliability of the results while making accurate predictions. The reliability  feedback of results can serve as a measure for future tasks. 
For instance, \emph{Yukun et al.}~\cite{medical_3} suggested that challenging cases with low confidence levels in the field of medicine should be reviewed by skilled surgeons.

In the past years, the research direction of confidence modeling has evolved in two directions.
The first direction is to quantify the confidence of predicted results with diverse heuristic approaches.
For example,  \emph{DeVries et al.}~\cite{devries2018learning} enhanced the model’s prediction by adding a branch of calculating the confidence value, based on the original classification task. The confidence value is utilized to identify whether the input sample is an out-of-distribution~(OOD) sample. \emph{Hendrycks et al.}~\cite{hendrycks2016baseline} used the predicted softmax probability of the sample as the confidence estimation and detected OOD samples by selecting the samples with the minimum softmax probability values.
\emph{Kendall et al.}~\cite{kendall2017uncertainties} argued that the model uncertainty could be explained by inherent noise in the captured data with Bayesian approaches.  
On the other hand, the confidence calibration work has also received extensive attention recently. For instance, \emph{Guo~et~al.}\cite{calibration} analyzed the overconfidence reasons~(i.e., model capacity, batch normalization, and weight decay) of models based on deep neural networks and  gave some post-processing techniques~(e.g., temperature scaling, matrix, and vector scaling) to deal with these problems. 
\emph{Moon, Jooyoung et al.}\cite{loss_1} introduced the correctness ranking loss to ensure the credibility of the predicted probability, which defines the optimization objection that the confidence estimate for  the correctly predicted sample is greater than the confidence estimate for the incorrectly predicted sample. 
However, these confidence estimation methods cannot be directly applied to the task of cognitive diagnosis.
In this paper, we propose a novel calibration loss method that aims to optimize parameters, thereby ensuring the reliability of the predicted probabilities, which allows the model to maintain confidence in its output results.

\begin{figure}[!t]
\centering
\vspace{-4mm}
\subfigure[]{
\raisebox{1.6mm}{\includegraphics[width=3.6cm]{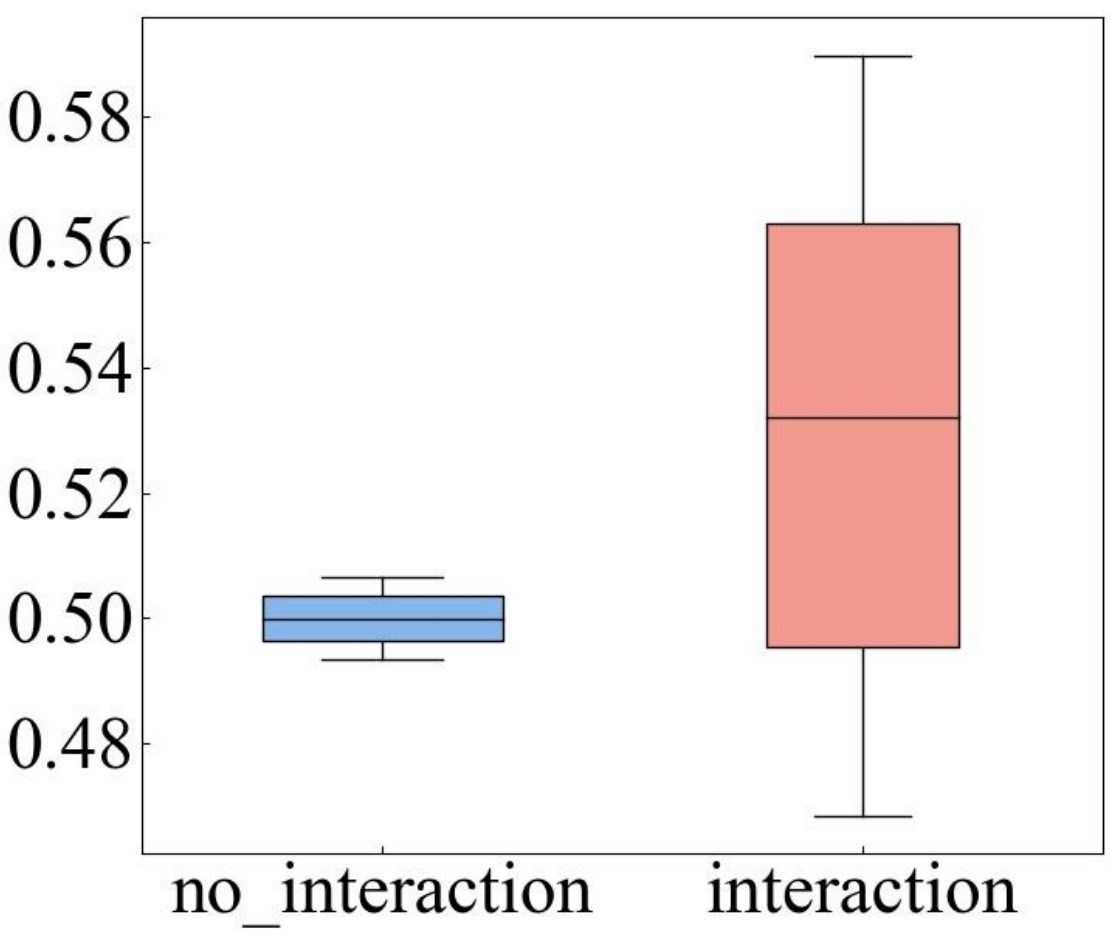}}
\label{preliminary_1}
}
\hspace{-1mm}
\subfigure[]{
\includegraphics[width=3.1cm]{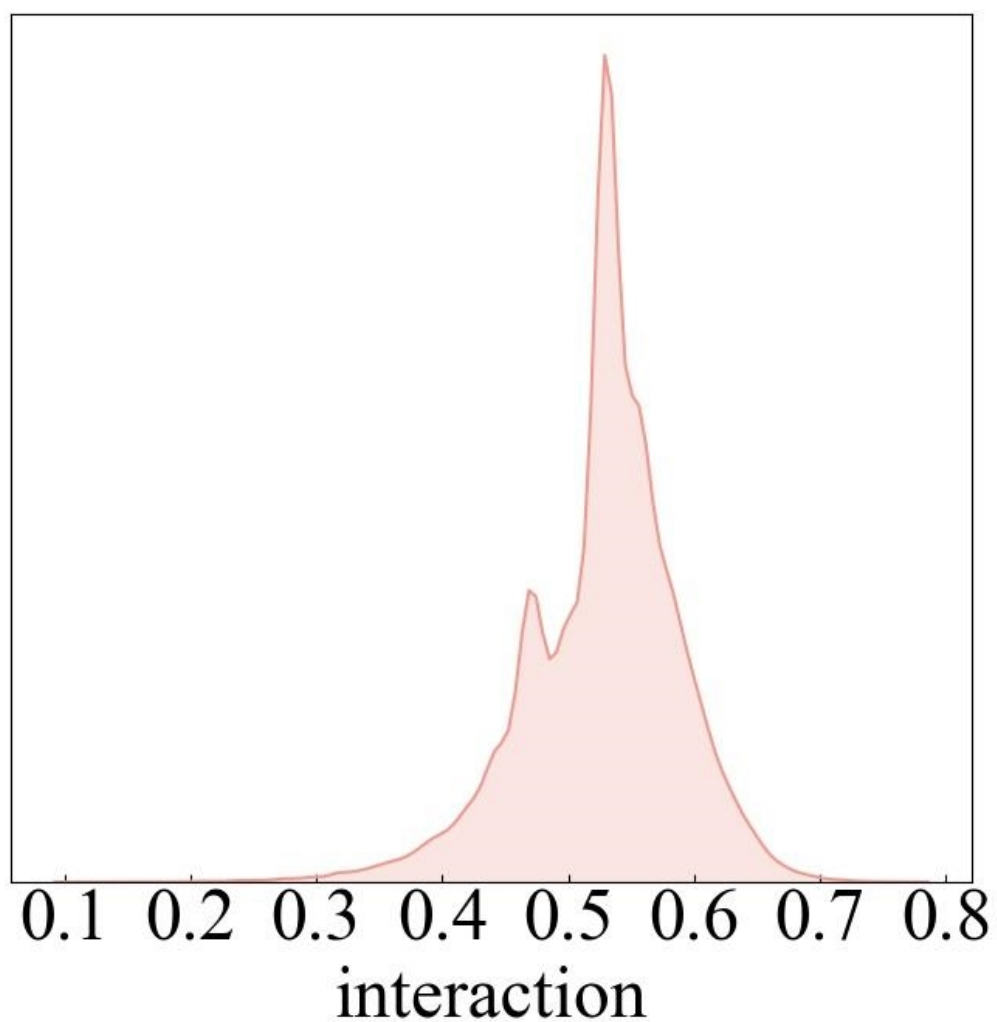}
\label{preliminary_2}
}
\vspace{-2mm}
\caption{(a)~The distribution of all students' ability status diagnosed by NCD on the Assist2009 dataset. The blue part represents diagnostic status of knowledge concepts not interacted with, and the red part represents diagnostic status of knowledge concepts interacted with. (b)~The density plot of all students' status on the knowledge concepts that they have interacted with.}
\vspace{-3mm}
\label{preliminary}
\end{figure}

\section{PRELIMINARIES}
In this section, we first introduce some currently known cognitive diagnosis functions~(i.e., IRT, MIRT, and NCD). Then, we analyze the diagnostic feedback of the previous CDMs using NCD as a case study.
Finally, we formally define the research problem being investigated in this paper.

\subsection{Cognitive Diagnostic Functions}\label{sec:cdf}

Generally, cognitive diagnosis in computational education aims to determine the student's ability status through the student exercising performance prediction task.

As a classic and representative diagnostic formula in educational psychology, IRT~\cite{IRT} portrays the student ability status of student $s_i$ with an integrated value $\theta_i \in R^1$. The logistic regression function is used to predict the probability $p(y_{ij} = 1)$ that the student $s_i$ will answer the exercise $e_j$ correctly as follows, 
\begin{equation} \label{eq1}
\begin{aligned}
p(y_{ij} = 1) = \frac{1}{1 + e^{-D\beta_j(\theta_i - \alpha_j)}},
\end{aligned}
\end{equation}
where $\alpha_j $ and $\beta_j \in R^1$ are difficulty and discrimination of exercise $e_j$ respectively. $D$ is a constant.

MIRT~\cite{MIRT} expands the student's ability status and exercise parameters from an integrated value to multi-dimensional vectors on the basis of IRT, so as to assess the student's ability status from multiple aspects. In this paper, we resemble some work~(e.g., RCD~\cite{RCD}) to map each dimension of MIRT to a specific knowledge concept by integrating the $Q$-matrix~\cite{NCD}. Under such consideration,
the probability that student $s_i$ with ability status vector $\theta_i$  makes a correct response to exercise $e_j$ with difficulty vector $\alpha_j$  can be expressed as: 
\begin{equation} \label{eq2}
\begin{aligned}
p(y_{ij} = 1) = \mathrm{\varrho}(f_{sum}(Q_j \circ (\theta_i - \alpha_j))),
\end{aligned}
\end{equation}
where $Q_j$ indicates which knowledge concepts are relevant to $e_j$, $\theta_i$ and $\alpha_j \in R^{K}$, $f_{sum}(.)$ is the sum operation and $\varrho(.)$ is the sigmoid function.

NCD~\cite{NCD} attempts to accommodate complex nonlinear interactions between students and exercises by building a new diagnostic function consisting of three fully connected layers~($f_{MLP}$)  and
one shallow layer inspired by MIRT. The cognitive diagnostic function of NCD can be formalized as:
\begin{equation} \label{eq3}
\begin{aligned}
p(y_{ij} = 1) = \mathrm{\varrho(}f_{MLP}(Q_j \circ (\theta_i - \alpha_j) \times \beta_j)),
\end{aligned}
\end{equation}
where $\theta_i$ is a $K$-dimensional vector representing the ability status of student $s_i$,
$\alpha_j$ and $\beta_j \in R^{K}$ are exercise difficulty and exercise discrimination, respectively.
The value of each dimension in $\theta_i$ indicates the student $s_i$' mastery level of the corresponding knowledge concept.

\begin{figure}[!t]
\centering
\vspace{-5.8mm} 
\subfigure[]{
\raisebox{1.3mm}{\includegraphics[width=3.65cm]{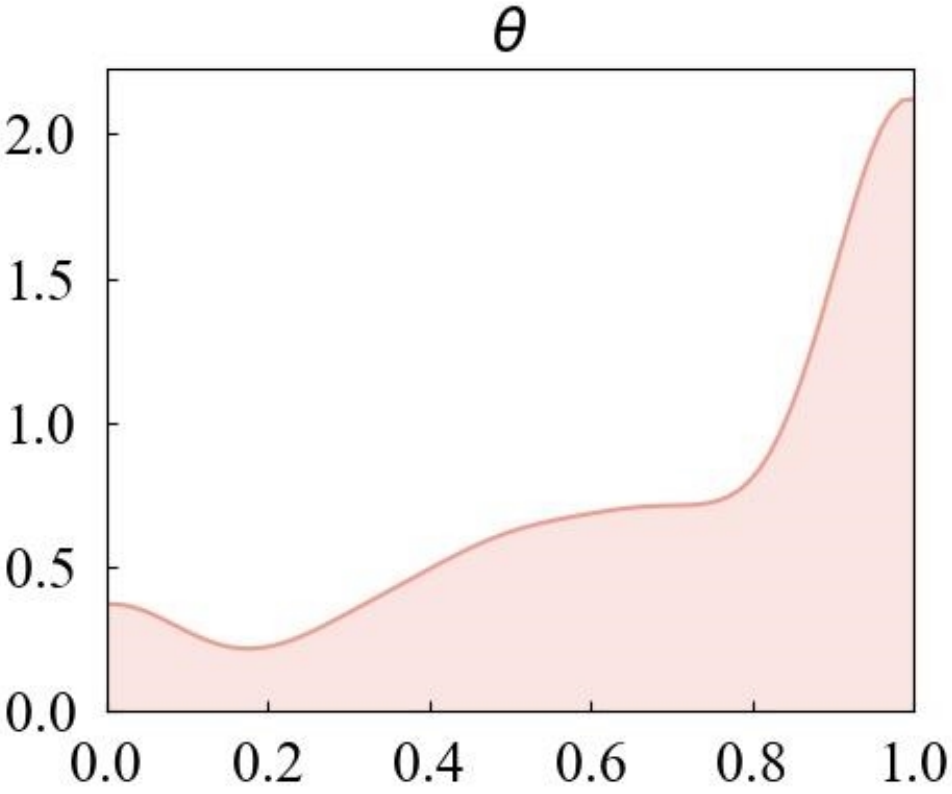}}
\label{preliminary_3}
}
\hspace{-1mm}
\subfigure[]{
\raisebox{1.3mm}{\includegraphics[width=3.7cm]{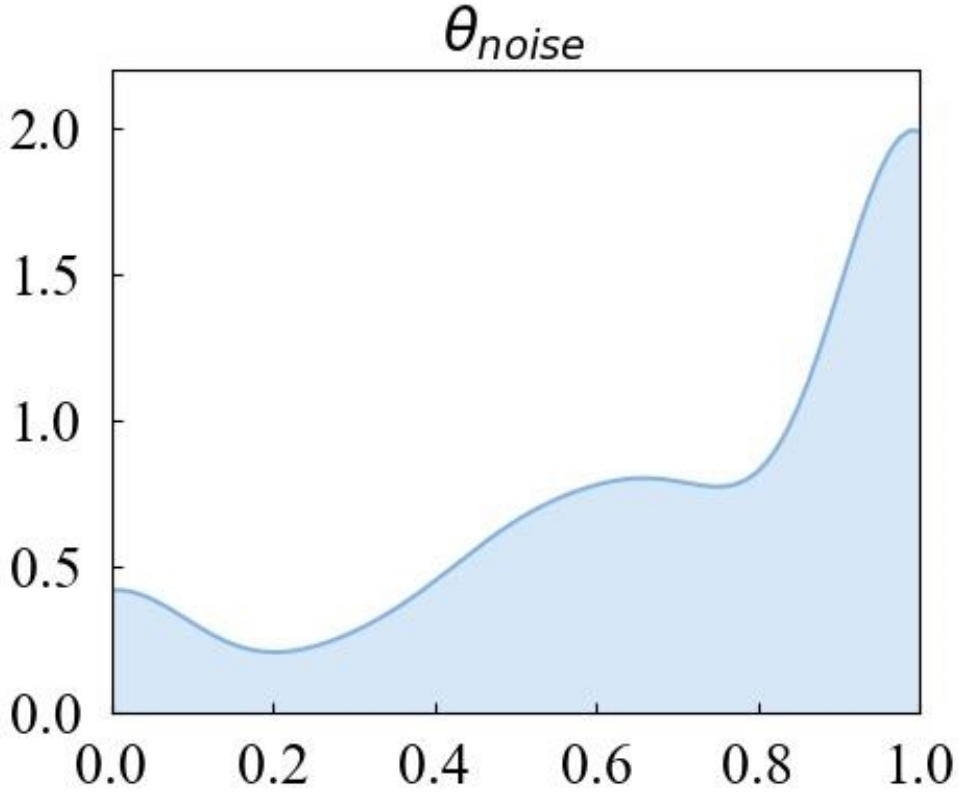}}
\label{preliminary_4}
}
\vspace{-2mm} 
\caption{(a) The density plot of the correct rate of students' performance prediction task related to knowledge concept \#50 by NCD in the test set of Assist2009. (b)~The density plot of the correct rate after randomly adding one noisy interaction data on concept \#50 for each student.}
\label{preliminary2}
\vspace{-3mm}
\end{figure}

\subsection{Diagnostic Feedback Analysis}
While current CDMs show remarkable accuracy in predicting student performance, we argue that their diagnostic feedback may not always be meaningful.

Without loss of generality, we take the NCD model as an example. Specifically, we trained the NCD on a public real-world dataset, namely Assist2009. Then, we can obtain all students' diagnostic feedback, i.e., their ability status $\theta_i$. As shown in Figure~\ref{preliminary}, we present the distribution of $\theta_i$. Here, the red part indicates all ability status $\theta_{il}$ for student $s_i$ on each knowledge concept $c_l$ that $s_i$ has interacted with it. Similarly, the blue part shows the ability status that the student has not interacted with.
Clearly, we can find that although the distributions of ability status values corresponding to the interactive knowledge concepts and non-interactive knowledge concepts are different, 
both of them have limited support included in [0.4, 0.6], which impedes the discriminate diagnostic feedback.

Moreover, we further analyze the impact of noisy interaction data on the diagnostic model. Here, Figure~\ref{preliminary_3} shows the density plot, which indicates the correct rate of students' performance prediction related to the knowledge concept \#50 in the test set based on the above NCD model on Assist2009. Next, we incorporate a randomly generated interaction for each student at knowledge concept \#50. The corresponding density plot on the correct rate of students' performance prediction is shown in Figure~\ref{preliminary_4}. We can find that the student's performance predictions were significantly degraded after incorporating the noisy data, which also demonstrates that even adding just one noisy interaction can undermine the reliability of diagnostic results.

Considering the aforementioned issues in the existing CDMs, in this paper, we focus on improving the reliability of diagnostic feedback by quantifying the confidence of the student's ability status. And the proposed framework, ReliCD, is designed to be adaptable to various cognitive diagnostic functions, including IRT, MIRT, and NCD.

\subsection{PROBLEM STATEMENT}

\subsubsection{Task Overview}
Cognitive diagnosis in intelligent education consists of three parts, a set of students $S=\{ s_1, s_2, ..., s_N  \}$, a set of exercises $E=\{ e_1, e_2, ..., e_M \}$ and  a set of knowledge concepts $C=\{ c_1, c_2, ..., c_K \}$, where $N$, $M$, $K$ represent the number of students, the number of exercises and the number of knowledge concepts, respectively. The relationship between exercises and knowledge concepts is represented by a $Q$-matrix predefined by experts, where the $Q$-matrix is defined as $\{Q\}_{M \times K}$. If exercise $e_j$ contains knowledge concept $c_l$, then $Q_{jl} = 1$, otherwise $Q_{jl} = 0$. The response logs $R$ include a set of triplets $<s_i,e_j,r_{ij}>$, where if the student $s_i$ answers exercise $e_j$ correctly, $r_{ij}=1$ otherwise $r_{ij}=0$. Along this line, we can formally the research problem in this paper as follows.

\textit{Problem Definition:} Given the students' answer logs $R$ and the experts' predefined $Q$-matrix, our goal is to diagnose the students' proficiency level for specific knowledge concepts and provide a confidence level for the diagnosis result.

\section{Method}
\begin{figure*}[t]
\centering
\vspace{-4mm}
\includegraphics[width=17cm]{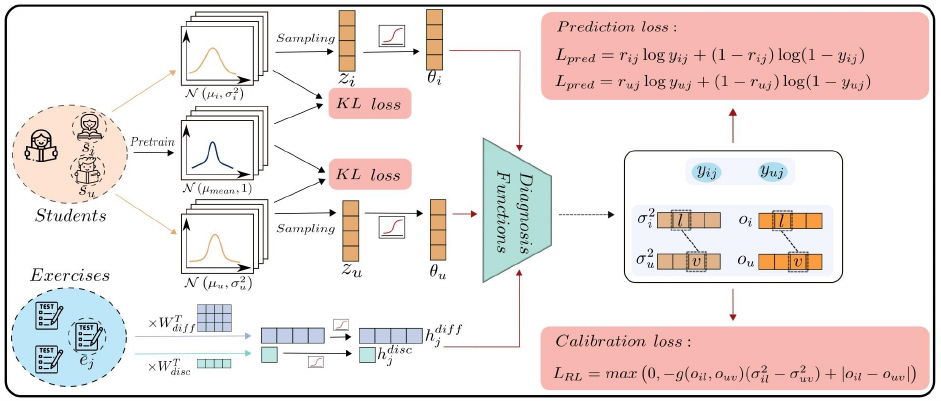}
\caption{The illustration of our basic idea in ReliCD. Each student $s_i$ is denoted by a personalized Gaussian distribution  variable  $z_i\sim \mathcal{N}(\mu_i,\sigma^2_i)$ and the corresponding ability state $\theta_i$ can be specified by applying the Sigmoid function on $z_i$, which is also a distribution with the support on $[0,1]$. Next, prior common cognition $\mathcal{N}(\mu_{mean}, 1)$ helps avoid the situation that students master all knowledge concepts in advance to 0.  Then, a calibration loss is induced to close the relationship between uncertainty and the reliability of the student's ability states by establishing a ranking relationship. }
\label{framework}
\end{figure*}

Cognitive diagnosis is the process of diagnosing the student's abilities $\theta$ in a particular skill or concept. However, the reliability of diagnosis can be affected by various factors, such as noise in the data and the sparsity of interaction data. To address this issue, it is crucial  to incorporate methods of modeling uncertainty in the diagnostic process, which encourages an accurate and reliable final diagnosis. Along this line, we design a novel reliable cognitive diagnosis~(ReliCD) framework. As shown in Figure~\ref{framework}, it can be divided into three parts: 1) the student’s state and uncertainty module, 2) the cognitive diagnosis module, and 3) the training objective. Additionally, we have employed two effective strategies, namely prior consensus and uncertainty regularization, to enhance the performance of our framework.

\subsection{Student's State  and Uncertainty}
To model the uncertainty in the diagnostic process,
we argue that there should be a deviation in the ability representations of students diagnosed by traditional score prediction methods. This deviation is caused by errors that can occur when students interact with the exercises, which can lead to unreliable diagnostic results.
To address this issue, we model the student's ability representation as a Gaussian distribution.
The mean parameter represents the average ability status, while the variance provides criteria for reliable diagnostic results. If the variance of the distribution is small, the diagnosis tends to be more reliable.

To obtain a personalized distribution representation~(Gaussian distribution) for each student $s_i$, we multiply the student's one-hot encoding by different transferable matrices to obtain the means and variance parameters, respectively, i.e.,
\begin{equation} \label{eq4}
\begin{aligned}
 q\left(z_i|x_i^s\right) =  \mathcal N \left(\mu_i, \sigma_i^2\right), 
\mu_i =  \mathbf{W}_{\mu}^T \times x_i^s , 
\log \sigma^2_i = \mathbf{W}_\sigma^T\times x_i^s ,
\end{aligned}
\end{equation}
where $\mu_i$ and $ \sigma^2_i \in R^{d}$ represent mean and variance parameters for student $s_i$, respectively. $x_i^s \in R^N$ is the one-hot encoded representation of student $s_i$. 
$\mathbf{W_\mu}$ and $\mathbf{W_\sigma}$ $\in$ $R^{N \times d}$ are different transferable matrices.
$N$ notes the number of students and $d$ indicates the dimensionality of the hidden vector~(we will discuss the setting of $d$ in detail in Section~\ref{sec:cd}).




Unlike previous student ability modeling techniques, here,  we randomly sample students $s_i$ ability representations $\theta_i$ from the constructed  Gaussian distribution~$q(z_i|x_i^s)$. This approach simulates deviations in diagnostic results caused by potential noise in interactions between student $s_i$ and exercises $e_j$. The details are as follows,
\begin{equation} \label{eq5}
\begin{aligned}
\theta_i = \varrho(z_i), \ \  z_i \sim q\left(z_i|x_i^s\right),
\end{aligned}
\end{equation}
where $\theta_i$ $\in$ $R^{ d}$ denotes the ability representation of student $s_i$. $z_i$ is a vector randomly sampled from the constructed Gaussian distribution of $s_i$. $\varrho$ is a Sigmoid activation function to ensure that each dimension of $\theta_i$ is in $[0,1]$.

\subsection{Cognitive Diagnosis}\label{sec:cd}
After modeling the student's ability status with uncertainty, we turn to predict exercise performance with cognitive diagnosis functions $f_{cd}$ in Section~\ref{sec:cdf}. Specifically, we first extract diagnostic factors from exercise, i.e., exercise difficulty $h^{diff}$ and exercise discrimination $h^{disc}$, which are required in all cognitive diagnosis functions. 
The details are as follows:
\begin{equation} \label{eq6}
\begin{aligned}
h_j^{diff} =  \varrho ( \mathbf{W}_{diff}^T\times x_j^e ) , ~ 
h_j^{disc} = \varrho (\mathbf{W}_{disc}^T\times x_j^e) ,
\end{aligned}
\end{equation}
 where $x_j^e$ $\in$ $R^{M}$ denotes the one-hot representation of exercise $e_j$. $h_j^{diff}~\in~R^{ d}$ and $h_j^{disc}~\in~R^{ 1}$ are exercise difficulty and exercise discrimination of $e_j$, respectively. $\mathbf{W}_{diff} \in R^{M \times d}$ and $\mathbf{W}_{disc} \in R^{M \times 1}$ are two transferable matrices.
$M$ indicates the number of exercises and $d$ denotes the dimensionality of the hidden vector. 



Here, the predict probability $p(y_{ij}=1)$, indicating student $s_i$ answers correctly on exercise $e_j$, can be derived as follows:
\begin{equation} \label{eq7}
\begin{aligned}
p\left(y_{ij}=1\right) = f_{cd} \left(\theta_i, h_j^{diff}, h_j^{disc}\right),
\end{aligned}
\end{equation}
 where $f_{cd}$ denotes the cognitive diagnostic function. Please noted that our framework is flexible for various diagnostic functions. Here we further present three popular diagnostic functions, i.e., IRT, MIRT, and NCD, and specify detailed rules for them as follows:

\begin{itemize}
\item $\mathbf{IRT}$: As shown in Eq.~(\ref{eq1}), IRT models student ability~$\theta$, exercise difficulty $h^{diff}$ and exercise discrimination $h^{disc}$ as a one-dimensional continuous scalar. Therefore,  because of the definition of IRT itself, we set the hidden vector dimension d=1 in Eq.~(\ref{eq4}) and Eq.~(\ref{eq6}).

\item $\mathbf{MIRT}$: For MIRT, we firstly let $h^{disc} = 1$ in Eq.~(\ref{eq6}). Then, we uniformly map students' ability representation  $\theta$ in Eq.~(\ref{eq4}) and exercise difficulty $h^{diff}$ in Eq.~(\ref{eq6}) to the $K$ dimension (i.e., $d=K$) to model MIRT from the perspective of knowledge concepts.

\item $\mathbf{NCD}$:  For NCD, as shown in Eq.~(\ref{eq2}), it models students' ability $\theta$ in Eq.~(\ref{eq4}) and exercise difficulty  $h^{diff}$ in Eq.~(\ref{eq6}) from the dimension of knowledge concepts, i.e., $d=K$.
\end{itemize}





\subsection{Training Objective}
To optimize the parameters for obtaining the students' ability status, we maximize the likelihood $p(r_{ij}|x^s_i)$, which indicates the true probability that the student $s_i$ answers the exercise $e_j$. Specifically, we follow the literature~\cite{pagnoni2018conditional, shen2021joint, shen2018joint} and utilize the evidence lower bound as the training objective, which is tractable. Formally, we have
\begin{equation} \label{eq8}
\begin{aligned}
\log p_{\phi}&\left(r_{ij} | x_i\right)  \geq \int \log p\left(r_{ij} |  z_i\right) p\left(z_i | x_i\right) dz \\
 & = -\underbrace{KL\left(q_{\varphi} (z_i | x_i) | p_{\phi}(z_i)\right)}_{L_{KL}} + \underbrace{E_{q_{\varphi}} \left[\log p(r_{ij}|z_i)\right]}_{L_{pred}},
\end{aligned}
\end{equation}
where $p_{\phi}(z_i)$ is the prior distribution for the ability status of students. $q_{\varphi} (z_i | x_i)$ is the posterior distribution we constructed for the student $s_i$. 
$\log p(r_{ij}|z_i)$ measures the likelihood that students with ability status $\theta_i=\varrho(z_i)$ answers correctly on exercise $e_j$.
We follow the variational autoencoder (VAE)~\cite{kingma2013auto, qin2022towards, wang2020personalized} and leverage the sampling strategy to approximate $L_{pred}$ with one sample. Then, $L_{pred}$ can be specified as:
\begin{equation} \label{eq9}
\begin{aligned}
L_{pred} = \log p(r_{ij}|z_i) = r_{ij}\log y_{ij} + (1-r_{ij})\log(1 - y_{ij}).
\end{aligned}
\end{equation}
When assuming that prior distribution $p_{\phi}(z_i)$ of student $x_i$ satisfies the standard Gaussian distribution, $L_{KL}$ in the Eq.~(\ref{eq8}) can be calculated by,
\begin{equation} \label{eq10}
\begin{split}
\begin{aligned}
KL\left(q_{\varphi} (z_i | x_i) || \mathcal N(0, 1)\right) 
&= \sum_{k=1}^K \frac{1}{2} \left(\mu_{ik}^2 + \sigma_{ik}^2 - \ln \sigma_{ik}^2 - 1\right),
\end{aligned}
\end{split}
\end{equation}
where $\mu_{ik}$, $\sigma_{ik}^2$ are the mean and variance of student $s_i$ on knowledge concept $k$.


Furthermore, based on the cognitive diagnosis scenario, we can define the reliability of a student's diagnostic feedback on a specific knowledge concept as the probability of correctly predicting a student's performance on the corresponding knowledge. Formally, the diagnostic feedback reliability can be defined as follows:

\begin{definition}
Given a student $s_i$, a cognitive diagnosis model $f_{cd}(\cdot)$, and $s_i$'s diagnostic feedback $\theta_i$, the reliability of diagnostic feedback (i.e., $\theta_{il}$) on a specific knowledge concept $c_l$ based on $f_{cd}(\cdot)$ is the probability of correctly predicting $s_i$'s performance on $c_l$, i.e., $\prod_{j} p(y_{ij}=r_{ij}|\theta_{il})$, where $y_{ij}$ belong to all the response logs of student $s_i$ answered exercise $e_j$ which cover the concept $c_l$ (i.e., $Q_{jl}=1$).
\end{definition}

Since we aim to utilize the standard deviation $\sigma_i$ to assess the reliability of each student $s_i$'s diagnostic feedback on different knowledge concepts, we design a novel calibration loss as a regularization term for the training objective. Specifically, given $\sigma_{il}$ and $\sigma_{uv}$ as the standard variances of student $s_i$'s and $s_u$'s ability representations on the knowledge concept $c_l$ and $c_v$, respectively, if $\sigma_{ij} > \sigma_{uv}$, we can assume the reliability of $s_i$'s diagnostic feedback $\theta_{il}$ should smaller than $\theta_{uv}$. Formally, we have the following hypothesis for ranking confidence levels.
\begin{assumption} \label{eq11}
    Given $\sigma_{il}$ of student $s_i$ and $\sigma_{uv}$ of student $s_u$, we have the relationship: 
    \begin{equation}
    \small
    \begin{split}
    \begin{aligned}
        & \sigma_{il} >= \sigma_{uv}   \Leftrightarrow  \\ &  \prod_{\substack{<s_i,e_j,r_{ij}> \\ \in R_i, \ Q_{jl}=1}} p\left(y_{ij}=r_{ij}|\theta_{il}\right) <=      \prod_{\substack{<s_u,e_j,r_{uj}> \\ \in R_u, \ Q_{jv}=1}} p\left(y_{uj}=r_{uj}|\theta_{uv}\right).
    \end{aligned}
    \end{split}
    \end{equation}
\end{assumption}


Considering the probability $p(y_{ij}=r_{ij}|\theta_{il})$ is generally impractical to directly obtain, we follow the idea from \cite{toneva2018empirical} and \cite{loss_1} and hypothesis the probability of correctly predicting student $s_i$ performance on a specific knowledge concept is proportional to the frequency of correct predictions of $s_i$ on the triples $<s_i,e_j,r_{ij}>\in R_i$ where $Q_{jl}=1$ during the training process. Along this line, we design a calibration loss in a pairwise manner as follows,
\begin{equation} \label{eq12}
\begin{aligned}
L_{RL} = {\rm max}\left(0, -g(o_{il}, o_{uv})(\sigma_{il}^2 - \sigma_{uv}^2) + |o_{il} - o_{uv}| \right),
\end{aligned}
\end{equation}
where $o_{il}$ denotes the proportion of the frequency of correct predictions of $s_i$ on concept $c_l$ over the total number of prediction on such interactions $<s_i,e_j,r_{ij}>\in R_i, \ Q_{jk}=1$ during the training process. The $g(\cdot,\cdot)$ is defined as: 
\begin{equation}\label{eq13}
\begin{aligned}
g(o_{il}, o_{uv})=\left\{
\begin{aligned}
            1, & \quad \text{if}, \quad o_{il} > o_{uv} \\
            0, & \quad \text{if}, \quad o_{il} = o_{uv} \\
            -1, & \quad \text{otherwise.}
\end{aligned}
\right.
\end{aligned}
\end{equation}

Moreover, to reduce the training time cost, we sample the pair $(\sigma_{il}, \sigma_{uv})$ under the current mini-batch when optimizing Eq.~(\ref{eq8}). That is, given a mini-batch of the input interactions $\{<~s^b_1,e^b_1,r^b_1~>~,$ $<~s^b_2,e^b_2,r^b_2~>, ... , <~s^b_B,e^b_B,r^b_B~>\}$, we obtain the pair of standard deviations based on the sampled $i$-th and $j$-th~(Here $i$ and $j$ only represent the $i$-th and the $j$-th instance) training instance pair, where $<~s^b_i,e^b_i,r^b_i~>$ denotes the $i$-th training instance in the current mini-batch and $B$ denotes the size of mini-batch.

For IRT, since it models the students' ability representation as a one-dimensional continuous scalar from a macro perspective, we revise the Eq.~(\ref{eq12}) as follows, 
\begin{equation} \label{eq14}
\begin{aligned}
L_{RL}^{IRT} = {\rm max}\left(0, -g(o_i, o_u)(\sigma_{i}^2 - \sigma_{u}^2 ) + |o_{i} - o_{u}| \right),
\end{aligned}
\end{equation}
where $o_i$ represents the proportion of the frequency of correct predictions of student $s_i$ over the total number of predictions on such interactions $<s_i,e_j,r_{ij}>\in R_i$ and $\sigma_{i}$ is the standard deviation of student $s_i$'s ability representation.



Finally, the total loss function is defined as:
\begin{equation} \label{eq15}
\begin{aligned}
L = L_{pred} + \gamma * L_{KL} + \beta * L_{RL},
\end{aligned}
\end{equation}
where $\gamma$ and $\beta$ are introduced to balance different items. Particularly, we follow  the approach of $\beta$-VAE~\cite{burgess2018understanding} to weight $L_{KL}$ for enhancing performance.

\subsection{Prior Consensus and Uncertainty Regularization  }
Here, we propose two strategies to further improve our model: adjusting the prior distribution of the student's state and regularizing the range of uncertainty.

\begin{algorithm}[t]
  \caption{The training process of ReliCD.}
  \label{alg:Framwork}
  \begin{algorithmic}[1]
    \Require
      Students' response logs $R$ and $Q$ matrix.
    \Ensure
      Each student's ability status $\theta_i$ and variance $\sigma_i^2$.
    \State Pretrain ReliCD with Eq. (\ref{eq16}) (set $\beta=0$);
    \State $\mu_{\text{mean}} = \frac{\sum_{m=1}^{N} \mu_m}{N}$;
    \While {not convergence}
      \State Sample a mini-batch $<s_i,e_j,r_{ij}>$;
      \State Obtain $\mu_i$, $\sigma_i$ from Eq.~(\ref{eq4});
      \State ${\sigma_i}^2 \gets  g_{x}(\sigma_i^2 - \alpha) + \alpha$;
      \State Sample $z_i \sim \mathcal{N}(\mu_i, \sigma_i^2)$ and obtain $\theta_i$;
      \State Generate $y_{ij}$ based on Eq. (\ref{eq7});
      \State Sample $B$ pairs $(\sigma_{il}, \sigma_{uv})$ randomly in this mini-batch;
      \State Compute gradients based on loss functions Eq. (\ref{eq15});
      \State Update all parameters; 
    \EndWhile
    \State Return $\theta_i$ and $\sigma_i^2$ for each student $s_i$.
  \end{algorithmic}
  
\end{algorithm}

\subsubsection{Prior Consensus }
Due to the potential difference between knowledge concepts, such as in terms of difficulty and discrimination, we assume that the prior distribution of the student's status on each knowledge concept is different and individual. To model the individual prior and prevent information leakage, we only use the training set to pre-train our model by setting $\beta=0$. Then, we average the ability state vector, i.e., $\mu_m $ of all students as the mean parameter $\mu_{mean}$ of the prior distribution, i.e, 
\begin{equation} \label{eq16}
\begin{aligned}
\mu_{mean} = \frac{\sum_{m=1}^{N} \mu_m}{N},
\end{aligned}
\end{equation}
 which represents the prior common cognition of all knowledge concepts.
This method is helpful for us to understand the overall level of students in advance, and avoid the situation that students master all knowledge concepts in advance to 0. 
Then, we train our entire model with the prior distribution $\mathcal N(\mu_{mean}, 1)$, where the variance parameter is set as 1. Therefore, the new KL loss can be defined as:
\begin{equation} \label{eq17}
\begin{split}
\begin{aligned}
&{KL}\left(\mathcal N(\mu_i,  \sigma_i^2 ) || \mathcal N(\mu_{mean}, 1)\right) \\ =& \frac{1}{2} \sum_{l=1}^{K} \left((\mu_{il} -\mu_{mean,l} )^2 + \sigma_{il}^{2} -\log( \sigma_{il}^2) - 1\right).
\end{aligned}
\end{split}
\end{equation}

\subsubsection{Uncertainty Regularization}

At the same time, inspired by~\cite{ma2019mae,DU-VAE}, we also believe that the variance of the modeling should be within a reasonable range, neither too fluctuating nor too smooth. So we follow the idea in the literature \cite{DU-VAE} and apply dropout to the variance parameter for each student $i$, which discourages the large variance,
\begin{equation} \label{eq18}
\begin{aligned}
\hat \sigma^2_{il} = g_{x_il}\left(\sigma^2_{il} - \alpha \right) + \alpha,
\end{aligned}
\end{equation}
where $g_{x_il}$ is an independent random variable generated from a standard Bernoulli distribution.
$\alpha$ is our empirically defined value.
At this point, the distribution we construct for student $s_i$ can be rewritten as $ q \left(z_i|x_i\right) =  \mathcal N \left(\mu_i, \hat \sigma_i^2\right)$. 

After designing the technical details of our framework with two strategies for enhancing performance, we can train our framework with the training data following Algorithm~\ref{alg:Framwork}.

\section{Experiment}

In this section, we will provide a detailed description of the benchmark datasets, baselines, and experimental setup. The designed experiments aim to answer the following questions:

\begin{itemize}
\item $\mathbf{RQ1}$: How does  our framework perform compare to state-of-the-art cognitive diagnosis models?

\item $\mathbf{RQ2}$: Whether the specially designed parts of our framework effective?

\item $\mathbf{RQ3}$: How do the hyperparameters influence the effectiveness of our framework?

\item $\mathbf{RQ4}$: Whether our study can improve the reliability of cognitive diagnosis models?
\end{itemize}



        
    
        



        
    
        

\subsection{Dataset Description}
We validated the performance of our framework on four real-world datasets, which are three public datasets namely Assistments2009~(Assist2009)~$\footnote{https://sites.google.com/site/assistmentsdata/home/assistment2009-2010-data/skill-builder-data-2009-2010}$, Junyi~$\footnote{ https://www.kaggle.com/datasets/junyiacademy/learning-activity-public-dataset-by-junyi-academy?resource=download}$ and ENEM~$\footnote{https://github.com/godtn0/DP-MTL}$, and a private dataset namely e-Math.    
ASSISTments2009~(ASSISTments 2009-2010 ``skill builder'') is a public dataset collected by the assistant online tutoring systems in the 2009-2010 academic year. 
Junyi is a public dataset collected by the Khan Academy in  2012 year.  e-Math is a private dataset collected by a well-known electronic educational product, mainly containing math exercises and response records of primary and secondary school students.
ENEM is a  Brazilian students' college entrance examination.

Table~\ref{tab1} shows the basic information of the four datasets, including the number of students, the number of exercises, the number of knowledge concepts, the total number of answer logs, the average number of answer logs per student, and the average number of knowledge concepts contained in each exercise. Moreover, we uniformly filtered out students with less than $15$ response logs to guarantee that there is enough data for modeling each student for all datasets. Along this line, we obtained $2,493$ students, $17,671$ exercises, and $123$ knowledge concepts in Assist2009; $1,000$ students, $712$ exercises, and $39$ knowledge concepts in Junyi; $517$ students, $1,582$ exercises, and $61$ knowledge concepts in e-Math; $10,000$ students, $185$ exercises and $4$ knowledge concepts in ENEM.

We divided each dataset into training set, validation set, and test set by splitting each student's response records at a ratio of $70\%: 10\%: 20\%$. And, we trained our framework on the training set, tune the parameters of our framework on the validation set, and verify the performance of our framework on the test set.

\begin{table}[t]
	\hspace{-2cm} 
	\centering
	\caption{The statistics of datasets.}
	\begin{tabular}{@{}lccccc@{}}
		\toprule
		\noalign{\smallskip}	
		Dataset & Assist2009  &Junyi  &e-Math & ENEM \\
		\noalign{\smallskip}\hline\noalign{\smallskip}
		\# Students & 4.1k   &1.0k & 1.9k &10k\\
		\# Exercises & 17.7k  &0.7k  & 1.6k &0.1k \\
		\# Knowledge concepts  & 123 &39  & 61 & 4 \\
		\# Response logs & 324k  &203k  &62k & 18500k \\
		\# Avg logs per student & 77.96 &203.94 &120.71 &185  \\
		\# Avg concepts per exercise & 1.19  &1.00 & 1.21 &1.00 \\	
		\noalign{\smallskip}
		\bottomrule
	\end{tabular} 
	\label{tab1}
\end{table}

\begin{table*}[htbp]
\vspace{-5mm}
\centering
\caption{Quantitative results on students' score prediction.}
\scalebox{0.9}{
\begin{tabular}{c c | c c | c c | c c}
\toprule
\textbf{Datasets} & \textbf{Metrics} & \textbf{IRT} & \textbf{Reli-IRT} & \textbf{MIRT} &\textbf{Reli-MIRT} & \textbf{NCD} & \textbf{Reli-NCD}
\\
\midrule
 \multirow{6}{*}{Assist2009} & ACC~(\% $\uparrow$) & 68.17 $\pm$ 0.06 & \textbf{69.56 $\pm$ 0.29}$\ast$ & 70.62 $\pm$ 0.43 & \textbf{71.12 $\pm$ 0.23} & 72.20 $\pm$ 1.11 & \textbf{72.55 $\pm$ 0.20}
\\
~& RMSE~($\downarrow$)  & 0.4554 $\pm$ 0.0054  & \textbf{0.4429 $\pm$ 0.0012}  & 0.4536 $\pm$ 0.0018  & \textbf{0.4478 $\pm$ 0.0007} & 0.4347 $\pm$ 0.0028 & \textbf{0.4311 $\pm$ 0.0010}
\\
~& AUC~(\% $\uparrow$)  & 69.15 $\pm$ 1.35 & \textbf{72.36 $\pm$ 0.12}$\ast$ & \textbf{72.53 $\pm$ 0.73} & 72.14 $\pm$ 0.09 & 75.10 $\pm$ 0.14 & \textbf{75.10 $\pm$ 0.32}
\\
~& ECE~(\% $\downarrow$)  & 4.75 $\pm$ 0.03 & \textbf{3.13 $\pm$ 0.15}$\ast$ & 9.97 $\pm$ 1.16 & \textbf{7.81 $\pm$ 0.23}$\ast$ & 6.97 $\pm$ 0.50 & \textbf{1.69 $\pm$ 0.19}$\ast$
\\
~& MCE~(\% $\downarrow$)  & 10.91 $\pm$ 0.39 & \textbf{10.58 $\pm$ 0.26} & 13.11 $\pm$ 1.28 & \textbf{12.21 $\pm$ 0.52} & 9.00 $\pm$ 0.19   & \textbf{3.85 $\pm$ 0.75}$\ast$
\\
\midrule
\multirow{6}{*}{e-Math} & ACC~(\% $\uparrow$) & 67.57 $\pm$ 0.41 & \textbf{70.00 $\pm$ 0.05}$\ast$ &  67.49 $\pm$ 0.42 & \textbf{69.20 $\pm$ 0.42}$\ast$ & 69.11 $\pm$ 0.32 & \textbf{69.13  $\pm$ 0.34}
\\
~& RMSE~($\downarrow$)  & 0.4564 $\pm$ 0.0014 & \textbf{0.4390 $\pm$ 0.0008} & 0.4595 $\pm$ 0.0022 & \textbf{0.4506 $\pm$ 0.0007} & 0.4427 $\pm$ 0.0030 & 
\textbf{0.4399 $\pm$ 0.0013}
\\
~& AUC~(\% $\uparrow$)  & 69.77 $\pm$ 0.36 & \textbf{74.20 $\pm$ 0.29}$\ast$ & 71.23 $\pm$ 0.24 & \textbf{72.52 $\pm$ 0.23}$\ast$ & 73.79 $\pm$ 0.21 & \textbf{74.12 $\pm$ 0.20}
\\
~& ECE~(\% $\downarrow$)  & 3.56 $\pm$ 0.37 & \textbf{3.14 $\pm$ 0.24} & 8.88 $\pm$ 0.42 & \textbf{5.56 $\pm$ 0.39}$\ast$ & 4.53 $\pm$ 0.01 & \textbf{1.17 $\pm$ 0.07}$\ast$
\\
~& MCE~(\% $\downarrow$)  & 10.37 $\pm$ 0.78 & \textbf{9.85 $\pm$ 1.08} & 13.60 $\pm$ 0.63 & \textbf{13.49 $\pm$ 0.34} & 5.90 $\pm$ 1.01 & \textbf{2.06 $\pm$ 0.01}$\ast$
\\
\midrule
 \multirow{6}{*}{Junyi} & ACC~(\% $\uparrow$) & 71.56 $\pm$ 0.27 & \textbf{75.31 $\pm$ 0.19}$\ast$ & 75.73 $\pm$ 0.17 & \textbf{75.79 $\pm$ 0.15} & 75.60 $\pm$ 0.26 & \textbf{76.05 $\pm$ 0.15}
\\
~& RMSE~($\downarrow$)  & 0.4342 $\pm$ 0.0021 & \textbf{0.4081 $\pm$ 0.0009}$\ast$ & 0.4291 $\pm$ 0.0004  & \textbf{0.4279 $\pm$ 0.0005} & 0.4068 $\pm$ 0.0006 & \textbf{0.4042 $\pm$ 0.0005}
\\
~& AUC~(\% $\uparrow$)  & 74.09 $\pm$ 0.39 & \textbf{78.84 $\pm$ 0.15}$\ast$ & 77.18 $\pm$ 0.11 & \textbf{77.34 $\pm$ 0.17} & 79.87 $\pm$ 0.13 & \textbf{80.11 $\pm$ 0.14}$\ast$
\\
~& ECE~(\% $\downarrow$)  & 3.89 $\pm$ 0.19 & \textbf{2.36 $\pm$ 0.12}$\ast$ & 10.68 $\pm$ 0.17 & \textbf{10.13 $\pm$ 0.13} & 1.97 $\pm$ 0.72 & \textbf{1.68 $\pm$ 0.88}
\\
~& MCE~(\% $\downarrow$)  & 8.45 $\pm$ 0.36 & \textbf{4.85 $\pm$ 0.26}$\ast$ & 20.90 $\pm$ 14.54 & \textbf{14.25 $\pm$ 0.17}$\ast$ & 3.07 $\pm$ 0.94 & \textbf{2.59 $\pm$ 1.04}
\\
\midrule
 \multirow{6}{*}{ENEM} & ACC~(\% $\uparrow$) & 71.70 $\pm$ 0.26 & \textbf{73.09 $\pm$ 0.44}$\ast$  & 70.91 $\pm$ 0.02 & \textbf{72.02 $\pm$ 0.03}$\ast$ & 73.45 $\pm$ 0.16 & \textbf{73.46 $\pm$ 0.12}
\\
~& RMSE~($\downarrow$)  & 0.4448 $\pm$ 0.0009 & \textbf{0.4319 $\pm$ 0.020} & 0.4514 $\pm$ 0.0002 & \textbf{0.4443 $\pm$ 0.0001}$\ast$ & 0.4288 $\pm$ 0.0008 &\textbf{ 0.4286 $\pm$ 0.0005}
\\ 
~& AUC~(\% $\uparrow$)  & 69.31 $\pm$ 0.14 & \textbf{72.18 $\pm$ 0.06}$\ast$ & 69.86 $\pm$ 0.08 & \textbf{69.92 $\pm$ 0.02} & 72.93 $\pm$ 0.10 & \textbf{72.96 $\pm$ 0.06}
\\
~& ECE~(\% $\downarrow$)  & 5.03 $\pm$ 0.15 &  \textbf{2.10 $\pm$ 0.09}$\ast$ & 7.78 $\pm$ 0.06 & \textbf{6.63 $\pm$ 0.09}$\ast$ & 1.64 $\pm$ 0.16 & \textbf{ 0.76 $\pm$ 0.08}$\ast$
\\
~& MCE~(\% $\downarrow$)  & 10.96 $\pm$ 0.46 & \textbf{3.71 $\pm$ 0.14}$\ast$   & 12.79 $\pm$ 0.11 & \textbf{7.56 $\pm$ 0.12}$\ast$ & 3.82 $\pm$ 0.19 & \textbf{1.64 $\pm$ 0.16}$\ast$
\\
\bottomrule
\end{tabular}} 
\label{tab2}
\vspace{-3mm}
\end{table*}


\subsection{Experimental Setup}
\subsubsection{Experimental settings}
In the experiment, we used Xavier initialization to initialize all parameters in our framework. 
We leveraged the Adam optimizer to train our reliable CDMs with a batch size of $32$ and a learning rate of $0.002$, respectively.
We used five-fold cross-validation to more accurately evaluate the performance of our framework on all datasets. As mentioned in Section 4.3, we set $\beta=0$ during the model pre-training phase. While during the training, validation, and testing phases, we set $\gamma$=1e-4, $\beta$=0.1. Our framework~\footnote{https://github.com/BIMK/Intelligent-Education/tree/main/ReliCD} and baselines were implemented with  Pytorch=$1.7.1$ by Python=$3.6$, and all experiments were conducted on an NVIDIA GeForce RTX 3090-24GHB.


\subsubsection{Evaluation metrics}
Here we evaluate our work from two aspects. The first aspect is the performance of our framework, which can be measured by ACC (Accuracy), RMSE (Root Mean Square Error), and AUC (Area Under an
ROC Curve), using the same metrics as previous work (e.g., NCD). 
The second is the quality of confidence estimation on the student's ability status, which can not be evaluated directly. Here, we turn to measure the confidence of our framework in predicting exercise performance by the expected calibration error~(ECE)~\cite{naeini2015obtaining} and the Maximum Calibration Error~(MCE)~\cite{naeini2015obtaining}, which are widely used in confidence estimation related literature~\cite{calibration,loss_1}.
The smaller the value of ECE or MCE, the better the quality of confidence estimation.
Specifically, we first divide the prediction probability interval into a certain number of bins. Then, ECE and MCE can be calculated by adding up and taking the maximum of the differences between the mean probability in each bin and the accuracy among the corresponding samples with weight, respectively.
The calculation formulas are as follows,
\begin{equation} \label{eq19}
\begin{aligned}
{\rm ECE} = \sum_{n = 1}^{M} \frac{|B_n|}{a} |{\rm acc}\left(B_n\right) - {\rm avgProb}\left(B_n\right)|,\\
{\rm MCE} = max_{n \in \left\{ 1, 2, ..., M\right\}} |{\rm acc}\left(B_n\right) - {\rm avgProb\left(B_n\right)}|,
\end{aligned}
\end{equation}
 where $M$ is the number of interval bins, $B_n$ denotes the set of samples with prediction probability in [$\frac{n-1}{M}$, $\frac{n}{M}$], $a$ is the total number of samples, ${\rm acc}(B_n)$ is the accuracy of the samples in $B_n$, ${\rm avgProb}(B_n)$ is the average prediction probability of our framework for the samples in $B_n$.









\begin{figure}[t]
\centering
\includegraphics[width=8cm]{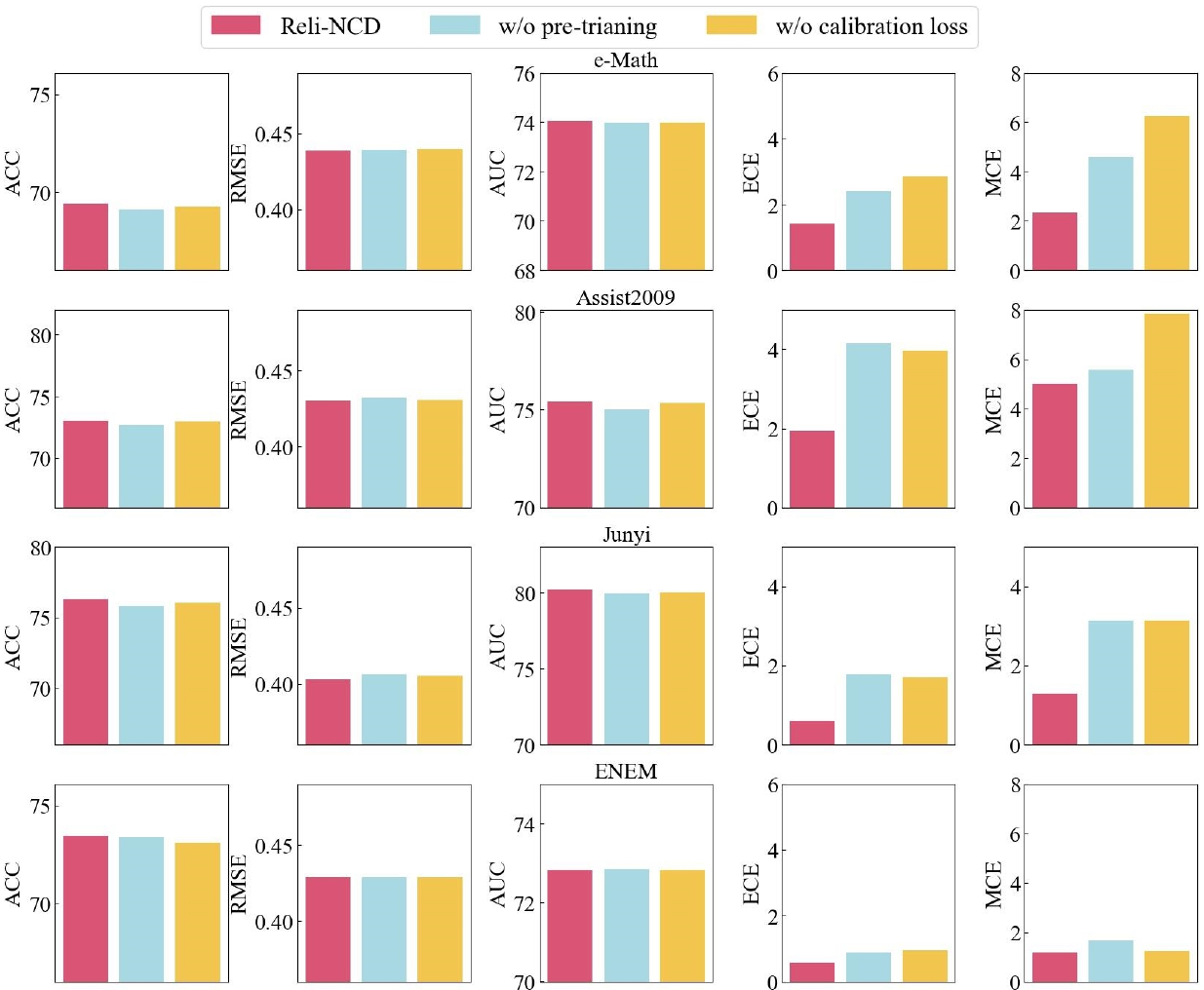}
\label{fig5_1}
\quad
\caption{Results of Reli-NCD and its variants.}
\label{fig3}
\vspace{-4mm}
\end{figure}

\subsection{Performance Comparison (RQ1)}

To verify the effectiveness of our proposed framework, we applied it to different cognitive diagnostic functions, including IRT, MIRT, and NCD. As a result, we obtained three reliable cognitive diagnosis methods: Reli-IRT, Reli-MIRT, and Reli-NCD. Our goal is to substantially enhance confidence metrics (i.e., ECE and MCE) while making slight improvements to traditional metrics (i.e., ACC, RMSE, and AUC).
Specifically, all these generated reliable cognitive diagnosis methods sampled the students' ability state from a constructed distribution to model students' uncertainty. As illustrated in Table~\ref{tab2}, we compared our ReliCDs with the corresponding baselines on four real-world datasets and we bolded the best experimental results with black lines. 
Moreover, we conducted the standard student t-test for the pair of our ReliCDs and the baselines at all indicators. Results are summarized in in Table~\ref{tab2} with significant improvement ($p$-$value < 0.01$) denoted with an asterisk~($\ast$).
We can obtain the following observations. 
Firstly, we can find that our ReliCDs have a significant decline compared to the corresponding baseline in ECE and MCE on almost all datasets. It not only shows that capturing the uncertainty of students can calibrate the confidence value of the prediction results but also demonstrates our method largely improves the reliability of the diagnostic results.
Secondly, we observed that our ReliCDs have significantly improved the performance in terms of ACC, AUC, and RMSE in some datasets. It reveals that estimating the uncertainty of students' ability status on different knowledge concepts can enhance the effectiveness of the student performance prediction. 
Thirdly, we found that our Reli-NCD achieved the best performance on all datasets. Meanwhile, we observed that basic NCD did not achieve the best performance on the ECE at Assistments2009 and e-Math. It also shows that our solution brings a good reliability improvement to strengthen cognitive diagnostic functions like NCD.



\subsection{Ablation Study (RQ2)}

To verify the effectiveness of each specially designed component of our framework, we constructed two variants of our ReliCD by removing the corresponding components. Without loss of generality, we used Reli-NCD as the baseline, which is an implementation of our framework with the specific diagnostic function NCD. The variants are described as follows:

\begin{itemize}
\item $\mathbf{w/o pre-training}$: It is a variant of Reli-NCD by removing the pre-training process, so as to explore its impact on the experimental results.

\item $\mathbf{w/o calibration loss}$: It is a variant of Reli-NCD by removing the calibration loss, so as to explore its impact on the experimental results.
\end{itemize}


The results are summarized in Figure~\ref{fig3}. Clearly, we can find that Reli-NCD has achieved the best performance on all datasets. Meanwhile, we observe that both pre-training strategy and calibration loss can significantly improve the performance of ECE and MCE in almost all datasets. Specifically, the pre-training strategy effectively reduces about 26.4\%, 19.2\%, and 37.9\% on ECE at Assist2009, Junyi, and e-Math, respectively. Correspondingly, the calibration loss effectively reduces about 10.2\%, 7.2\%, and 24.8\% on ECE at the above datasets. It clearly demonstrates the effectiveness of those components in our framework, which also answers $\mathbf{RQ2}$.



\begin{figure}[t]
\centering
\subfigure
[e-Math] 
{\includegraphics[width=3.cm]{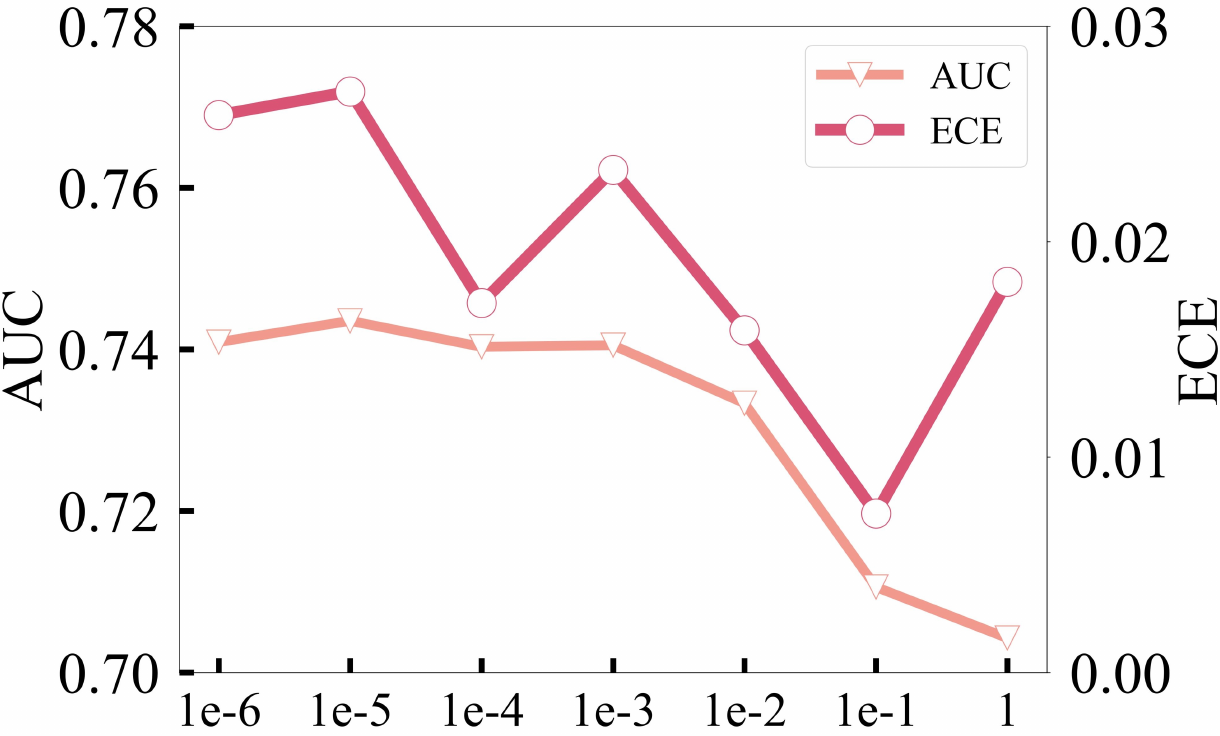}} 
\subfigure[Assist2009]
{\includegraphics[width=3.3cm]{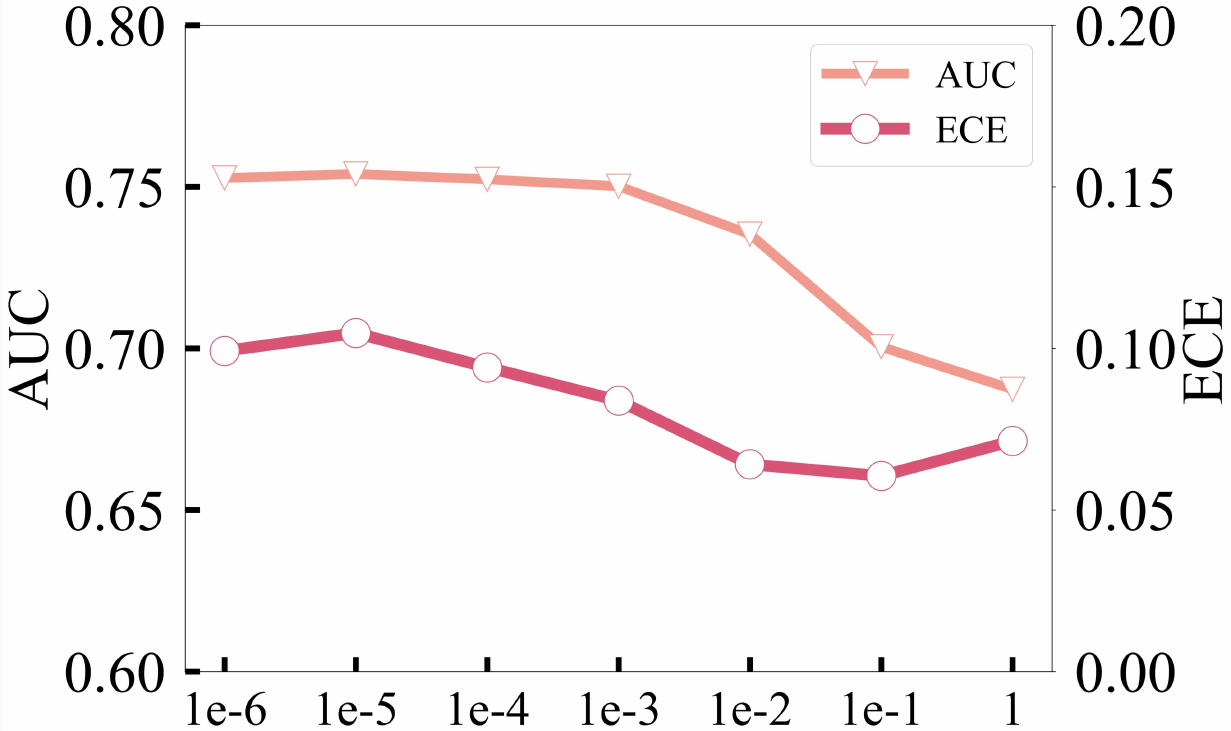}}
\subfigure[Junyi]
{\includegraphics[width=3.3cm]{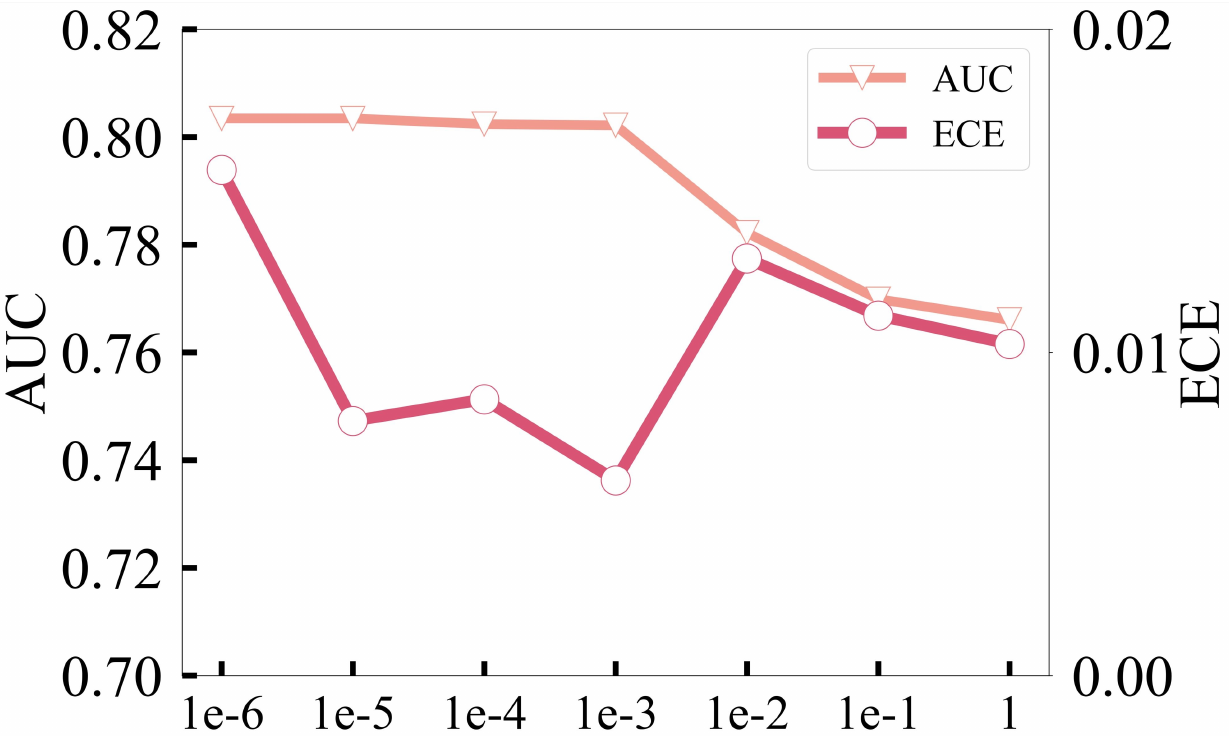}}
\subfigure[ENEM]
{\includegraphics[width=3.3cm]{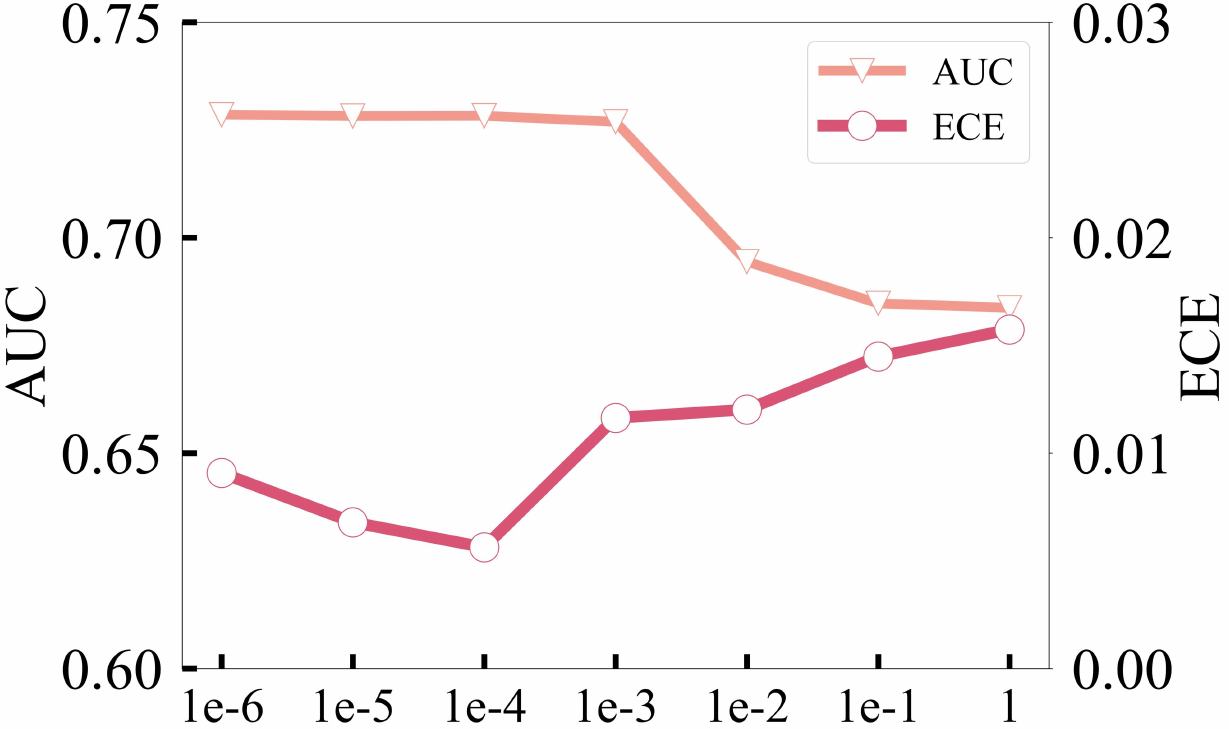}}
\caption{Impact of different sizes of $\gamma$ on the performance.}
\label{KL_analysis}  
\end{figure}

\begin{figure}[t]
\centering
\subfigure
[e-Math] 
{\includegraphics[width=3.3cm]{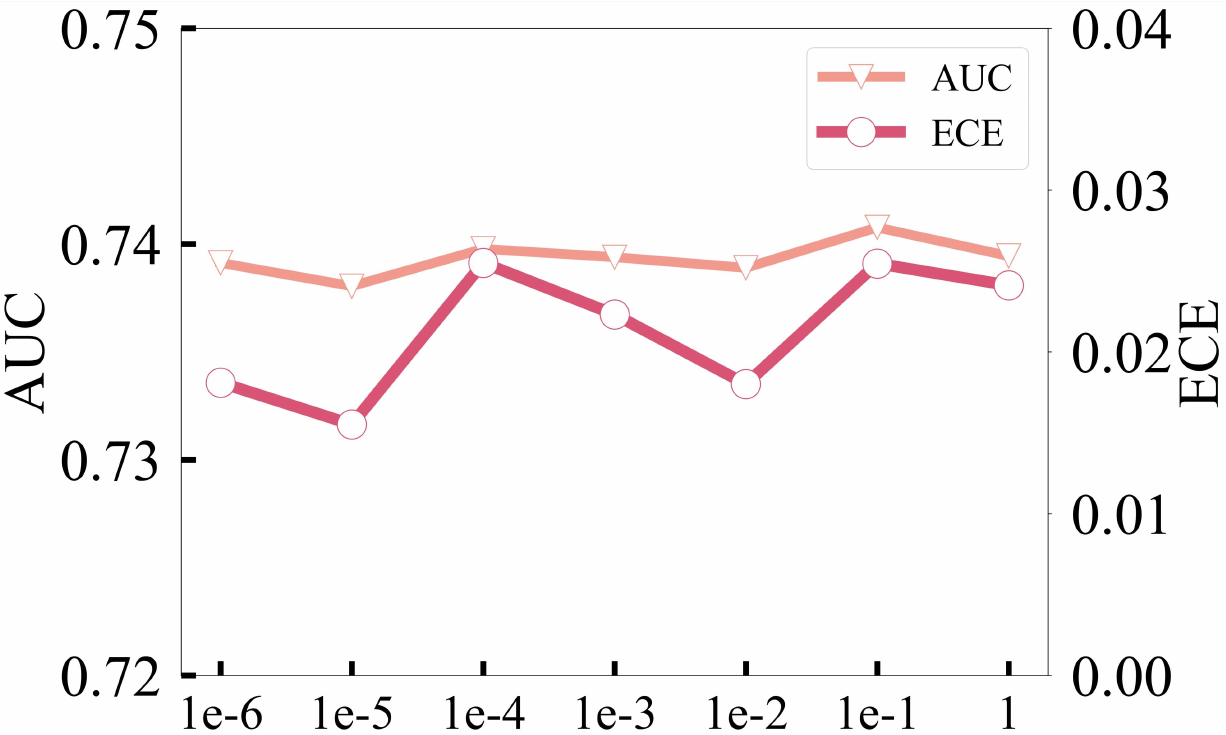}} 
\subfigure[Assist2009]
{\includegraphics[width=3.3cm]{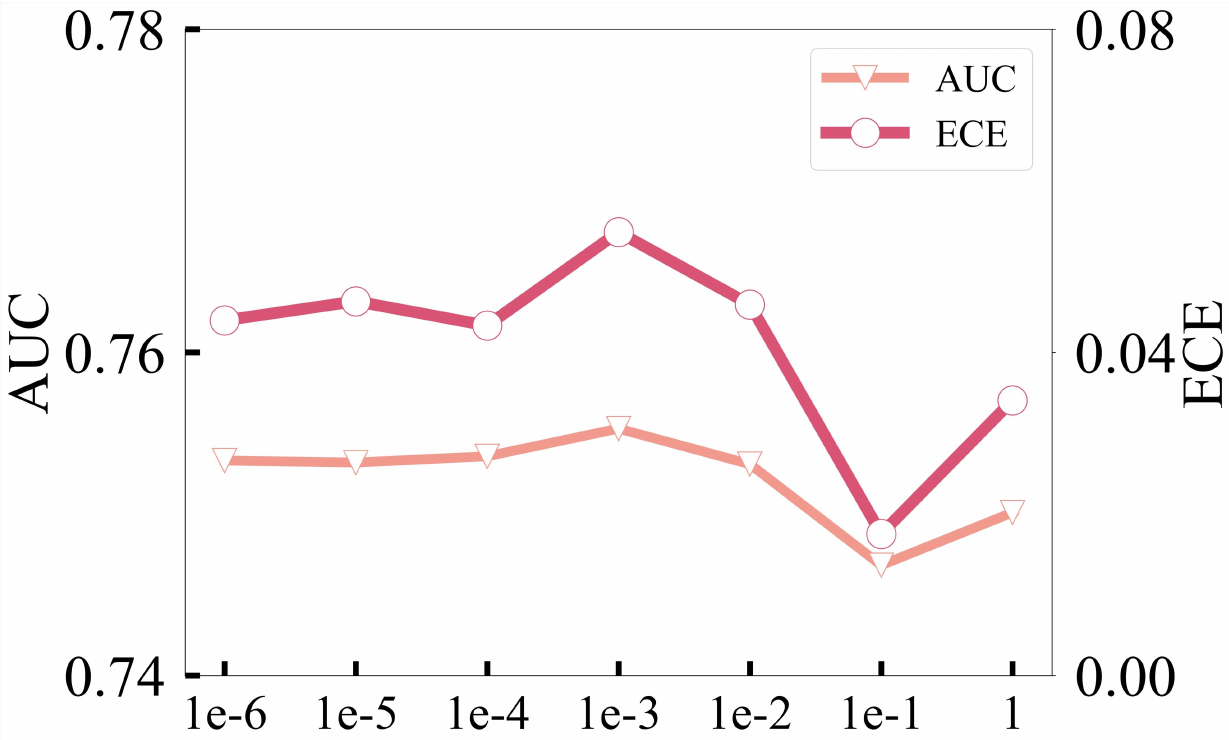}}
\subfigure[Junyi]
{\includegraphics[width=3.3cm]{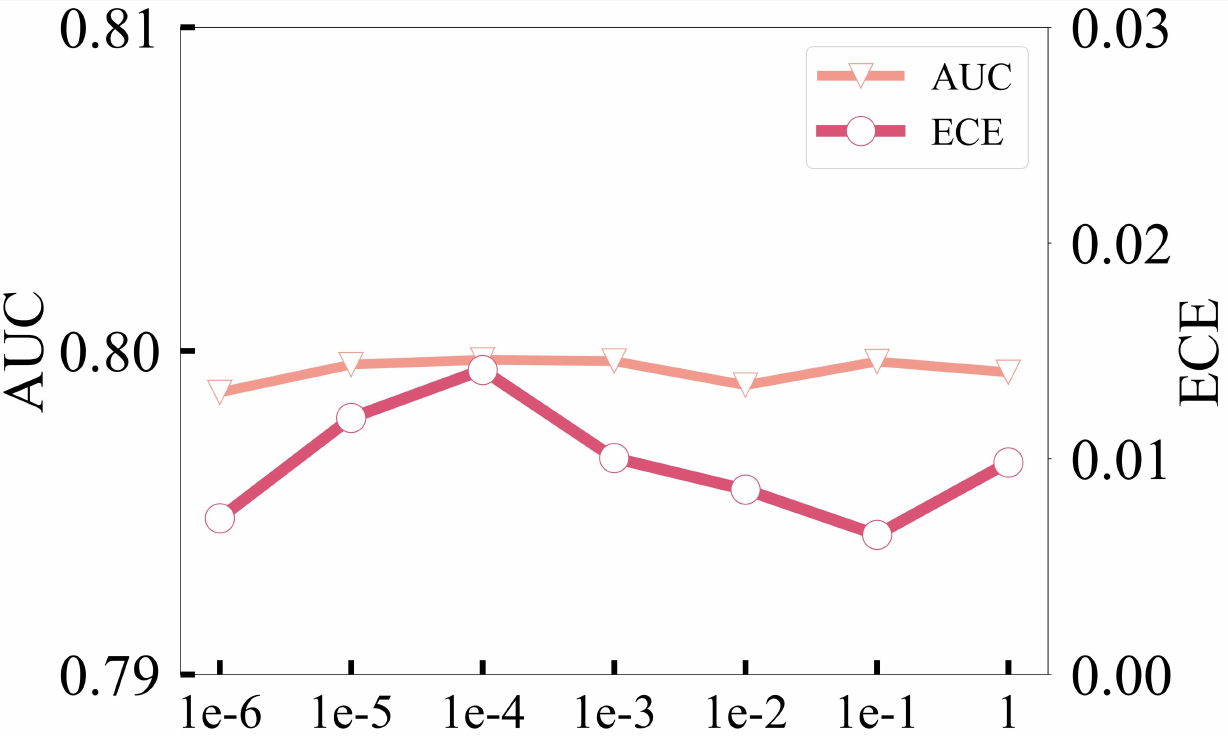}}
\subfigure[ENEM]
{\includegraphics[width=3.3cm]{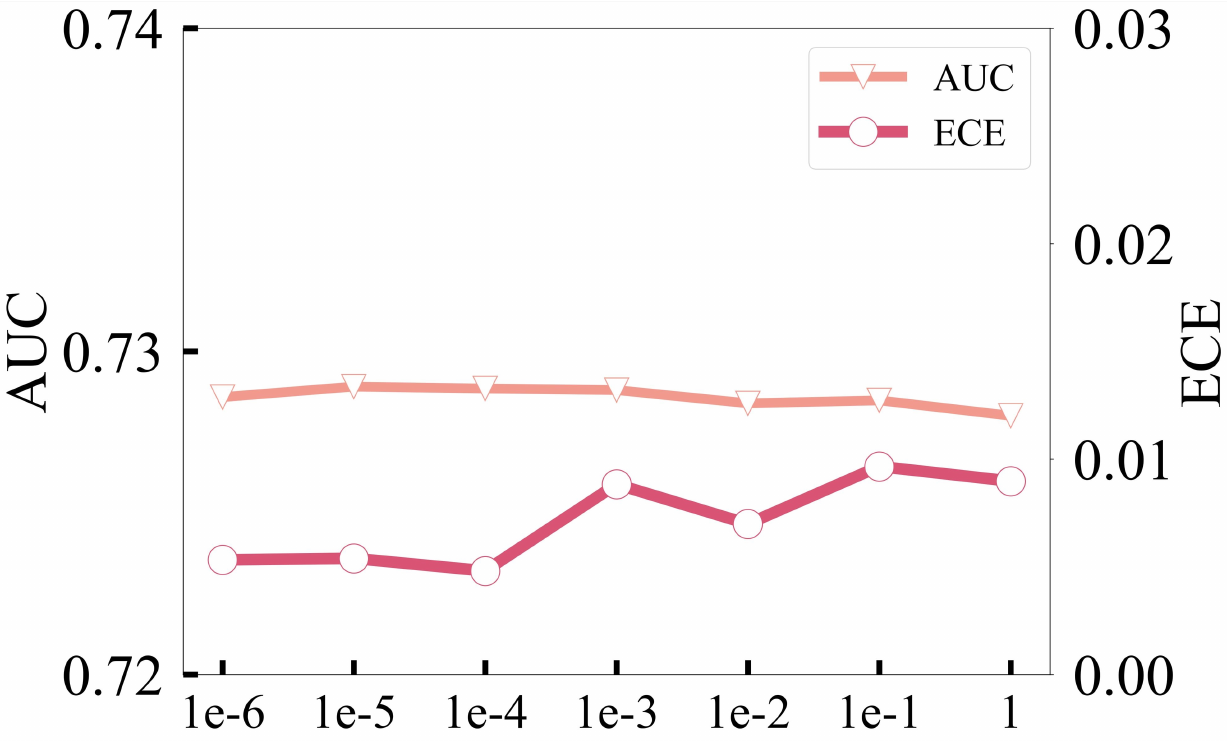}}
\caption{Impact of different sizes of $\beta$ on the performance.}
\label{ranking_analysis}  
\end{figure}



\subsection{Parameter Sensitivity Analysis (RQ3)}

To evaluate the sensitivity of hyperparameters $\gamma$ and $\beta$ in the loss function and answer the $\mathbf{RQ3}$, we conducted multiple experiments on e-Math, Assist2009, Junyi, and ENEM. We varied $\gamma$ and $\beta$ individually from $1e-6$ to $1$, while keeping the other parameter fixed.


As depicted in Figure~\ref{KL_analysis}, it is evident that the size of $\gamma$ has a significant effect on the results of both the AUC and ECE. The model's performance is optimal within the range of $0.0001-1$. This observation indicates that considering the constraints of student distribution within a reasonable range is beneficial to the model performance. As for $\beta$ shown in Figure~\ref{ranking_analysis}, varying the size of $\beta$ did not greatly affect the AUC values in e-Math and Assist2009, while the ECE values varied significantly on all three datasets under different sizes. The observation regarding $\beta$ suggests that the partial order relationship we established has a certain level of calibration effect on the final prediction of the model.

\begin{figure}[t]
\centering
\includegraphics[width=7cm]{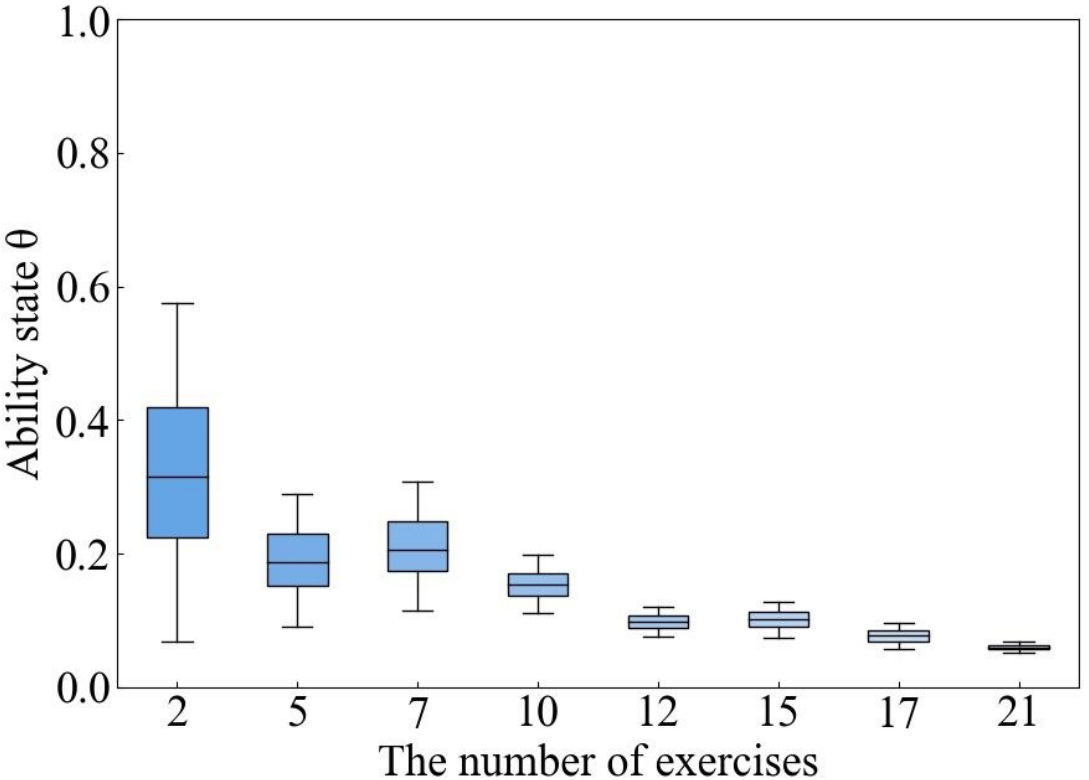}
\caption{The distribution of the student's ability statue $\theta$ under different numbers of interaction data.}
\label{case1}
\end{figure}

\begin{figure}[!t]
\centering
\subfigure[]{
\raisebox{1.3mm}{\includegraphics[width=3.6cm]{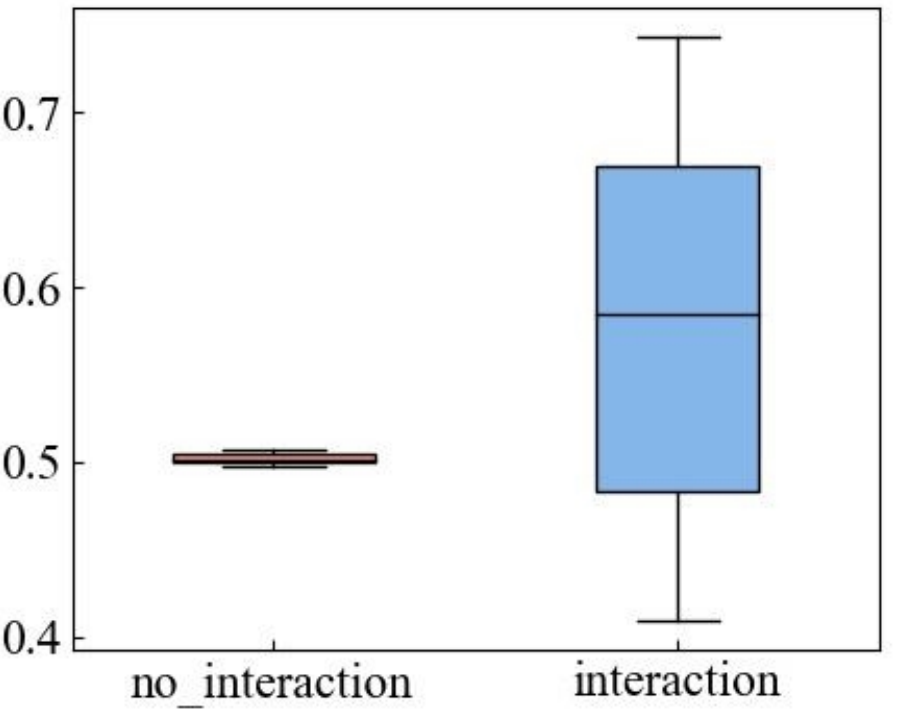}}
\label{case_2_1}
}
\subfigure[]{
\includegraphics[width=3.33cm]{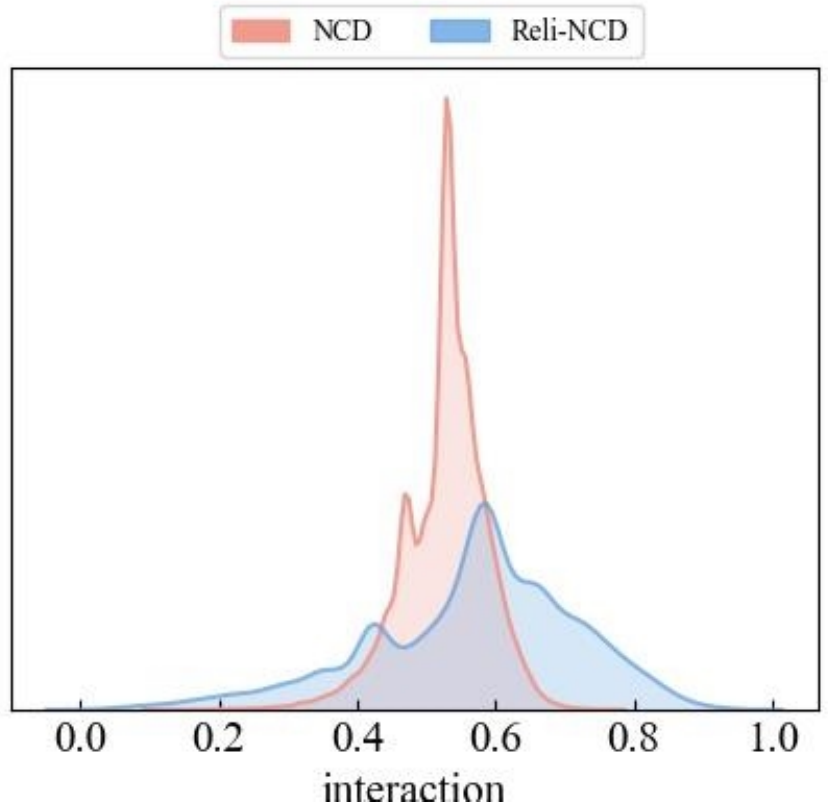}
\label{case_2_2}
}
\vspace{-2mm}
\caption{(a)~The distribution of all students' ability status diagnosed by Reli-NCD on the Assist2009 dataset includes the left part for knowledge concepts not interacted with and the right part for those interacted with. 
(b)~The density plot of all students' ability status of the knowledge concepts that they have interacted with. The red one is the predicted ability status based on NCD and the blue one is based on our Reli-NCD.}
\label{case_2}
\end{figure}




\subsection{Case Study (RQ4)}
Here we first show an example of predicted ability status via our framework. Specifically, we trained our Reli-NCD on the Assist2009 dataset. Figure~\ref{case1} shows the predicted student \#4164's ability status $\theta$ distribution of knowledge concept \#15 corresponding to training with different numbers of exercises on this concept. Clearly, we can observe that the fluctuation of student ability state decreases with more interaction data on this concept, while the mean value of the student ability status is also regionally stable. Therefore, our model can effectively identify the reliability of the diagnostic feedback, which will serve as a great aid to educators in assessing the usability of the diagnostic feedback. 

Moreover, similar to the diagnostic feedback analysis in the preliminaries, we trained our Reli-NCD on the Assist2009 and obtained the distributions of students' ability status of the interactive knowledge concepts and non-interactive knowledge concepts. As shown in Figure~\ref{case_2}, we can find that our Reli-NCD provides more discriminate diagnostic feedback than NCD, as its predicted ability status distribution span is significantly wider. Meanwhile, we obtained our Reli-NCD can achieve concentrated distribution on students' ability status of the knowledge concepts that they have not interacted with, which demonstrates the reliability of our diagnostic feedback.



\section{Conclusion}
In this paper, we introduced a reliable cognitive diagnosis framework with confidence awareness, namely ReliCD, which is the first one to quantify the confidence of the diagnosis
feedback and is flexible for different cognitive diagnostic functions. To be specific, we first proposed a Bayesian method to explicitly estimate the state uncertainty of different knowledge concepts for students, which enables the confidence quantification  of diagnostic feedback. In particular, to avoid the potential difference, we proposed to model the individual prior distribution for the latent variables of different ability concepts with a pre-trained model. Then, we introduced a logical hypothesis for ranking confidence levels. Moreover, we designed a novel calibration loss to optimize the confidence parameters by modeling the process of student performance prediction.  
Finally, we have conducted extensive experiments on $4$ real-world datasets, and the experimental results have clearly demonstrated the effectiveness of our ReliCD.
\section{Acknowledgments}
This work was supported in part by the National Natural Science Foundation of China under Grant 62107001, in part by the National Key Research and Development Project (NO. 2018AAA0100105), in part by  the Anhui Provincial Natural Science Foundation (NO. 2108085QF272 and No. 2208085QF194), in part by the University Synergy Innovation Program of Anhui Province under Grant GXXT-2021-004.

\bibliographystyle{IEEEtran}
\bibliography{refer}

\end{document}